\documentclass[fleqn,usenatbib]{mnras}

\usepackage{mathptmx}
\usepackage{comment}

\usepackage[T1]{fontenc}

\DeclareRobustCommand{\VAN}[3]{#2}
\let\VANthebibliography\thebibliography
\def\thebibliography{\DeclareRobustCommand{\VAN}[3]{##3}\VANthebibliography}

\usepackage{graphicx}	
\usepackage{amsmath}	
\usepackage{amssymb}	
\usepackage{xcolor}

\definecolor{ochre}{rgb}{0.8, 0.47, 0.13}

\newcommand{\ba}{\begin{eqnarray}}
\newcommand{\ea}{\end{eqnarray}}
\newcommand{\be}{\begin{equation}}
\newcommand{\ee}{\end{equation}}

\def\e1{e_1^2}

\newcommand{\orcid}[1]{\href{https://orcid.org/#1}{\includegraphics[width=8pt]{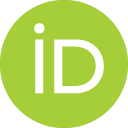}}}

\title[Migration Traps in AGN]{The Effect of Thermal Torques on AGN Disc Migration Traps and Gravitational Wave Populations}

\author[Grishin et al.]{
Evgeni Grishin ,$^{1,2}$ \orcid{0000-0001-7113-723X} \thanks{E-mail: evgeni.grishin@monash.edu} Shmuel Gilbaum$^3$ \orcid{0000-0002-6462-6657}, and Nicholas C. Stone$^3$\orcid{0000-0002-4337-9458}
\\
$^{1}$School of Physics and Astronomy, Monash University, Clayton, VIC 3800, Australia\\
$^{2}$The ARC Centre of Excellence for Gravitational Wave Discovery -- OzGrav, Clayton, VIC 3800, Australia\\
$^{3}$Racah Institute of Physics, The Hebrew University, Jerusalem, 91904, Israel\\
}

\begin{document}
\label{firstpage}
\pagerange{\pageref{firstpage}--\pageref{lastpage}}
\maketitle

\begin{abstract}
Accretion discs in active galactic nuclei (AGN) foster black hole (BH) formation, growth, and mergers. Stellar mass BHs 
migrate inwards under the influence of hydrodynamical torques unless they encounter a region where the torque flips sign.  At these migration traps, BHs accumulate and merge via dynamical or gas-assisted interactions, producing high-frequency LIGO/Virgo/KAGRA (LVK) gravitational wave (GW) sources and potentially cutting off the supply of extreme mass ratio inspirals that would otherwise make low-frequency, {\it LISA}-band GWs.  In this paper, we study the interplay between different types of migration torques, focusing especially on the ``thermal torques'' generated by the thermal response of the AGN to embedded stellar-mass BHs that accrete through their own mini-discs.  In contrast to previous work, we find that Type I torques cannot produce migration traps on their own, but thermal torques often do, particularly in low-mass AGN.  The migration traps produced by thermal torques exist at much larger distances ($\sim 10^{3-5}$ gravitational radii) than do previously identified Type I traps, carrying implications for GW populations at multiple frequencies.  Finally, we identify a bifurcation of AGN discs into two regimes: migration traps exist below a critical AGN luminosity, and do not at higher luminosities.  This critical luminosity is fit as $\log_{10} L_{\rm AGN}^c = 45 - 0.32 \log_{10}{(\alpha/0.01)}$ where $\alpha$ is the Shakura-Sunyaev viscosity parameter, a range compatible with recent claims that LVK GWs are not preferentially associated with high-luminosity AGN.
\end{abstract}

\begin{keywords}
galaxies-Active --  stars: Black Holes --   accretion discs -- gravitational waves
\end{keywords}




\section{Introduction} \label{1}

At present, following the end of the third (``O3'') observing run, the merger of binary black holes (BHs) has been seen over $90$ times by the LIGO-Virgo-KAGRA collaboration \citep{abbotto3b}. Upcoming observing runs are expected to significantly increase this number \citep{abbott18}, but the astrophysical origins of these mergers remain hotly debated. The most commonly studied channels for BH mergers are isolated binary \citep{belczynski16, neijssel19} and triple \citep{antonini17, ll19} evolution on the one hand, and dynamical, scattering-driven evolution in dense environments such as open \citep{ku19}, globular \citep{rod16,sam18}, and nuclear star clusters  \citep{antonini12, gri18, f19} on the other. For a comprehensive review and discussion of these two classic channels, see \cite{mandel22}.

In active galactic nuclei (AGN), the existence of a large-scale accretion disc around a massive black hole (MBH) creates fertile ground for binary BH (BBH) mergers, and these environments may be major contributors to the overall BBH merger rate \citep{bellovary16, bartos17, stone17, mckernan2018, tagawa2020, mckernan2020, samsing_nat, gen22}.  The AGN channel has a number of distinctive signatures that are (or may be) associated with it.  Direct electromagnetic counterparts caused by accretion variability are one \citep{mckernan2019, kimura2021, tagawa2023}, and a detection of this has even been claimed for GW190521\footnote{See also \citet{graham2023}.} \citep{graham2020}, although this association remains contested \citep{palmese2021} and, from a theoretical perspective, it is challenging for stellar-mass BHs to outshine the central MBH \citep{mckernan2019}.  As gravitational-wave (GW) samples grow, ``indirect'' electromagnetic counterparts (i.e. the statistical over-representation of AGN in sky error volumes) may be able to decisively constrain or measure the fraction of GWs originating in AGN \citep{bartos2017b, veronesi2022, ver23}.

Another potential signature of the AGN channel is unusually massive BH mergers: the same dense gas that hardens wide BH binaries may also grow them through accretion \citep{gilbaum2022}, and the deep gravitational potential well of the central MBH will retain the merger product after gravitational-wave recoil, producing hierarchical mergers \citep{mckernan2012} either through the accumulation of stellar-mass BHs in migration traps (\citealt{bellovary16, secunda19}; see below), or in dissipative single-single flybys \citep{tagawa2021, delaurentiis2023, li23, whitehead23, rowan23}, thus populating the ``upper mass gap'' forbidden for single-star evolution due to the onset of pair-instability \citep{farmer2019}. Indeed, the aforementioned GW190521 has a total mass of $ 142^{+28}_{-16} M_\odot$, with one progenitor BH massing $85^{+21}_{-14} M_\odot$  \citep{190521}, placing it firmly in the pair instability mass gap. 
It is worth noting that stellar clusters may initially have had gas that also contributed to their BH mass growth \citep{rozner22, rozner22_2}.

One of the key ingredients governing BHs embedded in AGN discs, and more generally in accretion disc physics, is orbital migration \citep[e.g.,][]{gt80, ward1997, tanaka2002}. In the classical, linear, picture, the gravitational interaction between a massive body and the nearby gas forms two overdense spiral arms. The angular velocity of the gas is slightly sub-Keplerian due to the typical negative pressure gradient in accretion discs, creating an asymmetry in the wakes' profile. This asymmetry prevents the cancellation of the two torques, and a smaller, usually negative net torque drives the massive body inward \citep[e.g.,][]{paardekooper2014}. This classical, ``Type I'' migration is usually the most relevant kind of hydrodynamic torque for embedded objects on prograde orbits, though, for more comparable mass ratios, the embedded BH can open a gap in the AGN disc and enter the (usually slower) ``Type II'' migration regime \citep{ward1997, crida2006}. Gaps 

While Type I torques are usually negative in sign (producing inward migration), they can sometimes change sign and produce a migration trap \citep{paardekooper2006, lyra2010}. 

Numerous recent developments have changed the classical picture of Type I migration (see \citealp{Paardekooper_review} for a recent review). Notably, \citet{jm17} have updated the Type I migration torque formulae by calibrating corotation and Lindblad torques from 3D hydrodynamical simulations; the results produce significant changes to the classic Type I formulae calibrated from 2D simulations that have seen wide use in the AGN literature \citep{paardekooper2010}.

Another recently quantified phenomenon in disc migration is the thermal torque. Thermal torques originate in discs with finite thermal diffusivity. In the absence of external heating, the gas near the vicinity of the massive body loses energy and has two cold and overdense regions \citep{lega14}. For a luminous body, the effect is opposite, and the released heat diffuses and forms two hot, underdense lobes \citep{b-l15}. Due to the offset from corotation, the contribution from these lobes is not symmetric and a net torque is generated. Recently, \cite{masset17} analysed both effects analytically using linear perturbation theory and derived 
a formula for both hot and cold thermal torques, which has been numerically validated by \cite{hankla2020}. 

Orbital migration has been extensively studied in the context of planet formation and dynamics  (see \citealp{Paardekooper_review} for  review and references therein)  and is thought to be responsible for resonant chains of close-in planets (\citealp{kley2004}), which may become unstable once the gas dissipates (\citealp{izidoro17}). The gravitational torques are manifested by detailed images of protoplanetary discs in star-forming regions by the Atacama-Large-Millimeter-Array \citep[ALMA]{alma15}. The kinetic signatures of gaps, spirals, and asymmetric and misaligned discs allow the detection of young planets (e.g., \citealp{pinte19}).

For AGN discs, such observations are not feasible and migration traps are hypothetical. Most of the theory of BH migration in AGN is borrowed from planetary dynamics. 
\cite{bellovary16} used the type I migration torque formula from numerical fits for protoplanetary discs \citep{tanaka2002, paardekooper2010} and applied it to the structure of an AGN disc which experiences a density/pressure maxima at $\sim 700$ gravitational radii \citep{sg03}, which subsequetly forms a migration trap. \cite{secunda19} showed that BH populations efficiently migrate into the trap and form resonant chains (similarly to resonant chains in multiplanet systems, \citealp{izidoro17}). The enhanced BH density in the migration trap is susceptible to binary formation and mergers \citep{tagawa2020, groebner2020}. \cite{hankla2020} numerically studied the effects of thermal torques in accretion discs and commented on their relevance in both protoplanetary discs and AGN discs alike.  \cite{hankla2020} derived a criterion for which thermal torques will overcome the type I torques, but focused on the luminous torques alone and not on the cold thermal torques. Importantly, a broad study of thermal torques in AGN disks with different parameters has not yet been performed, leaving the ultimate importance of this novel torque unclear.

The implications of migration in AGN discs go beyond high-frequency, LVK-band GWs.  At the $\sim$mHz frequencies that will be probed by future space-based interferometers such as {\it LISA} \citep{amaroseoane23}, one of the primary sources is the extreme mass ratio inspiral, or EMRI \citep{amaroseaone07, babak17}: a long-lived GW signal produced as a stellar-mass compact object gently inspirals into a MBH \citep{ryan95, sigurdsson97}.  As with LVK-band BBH mergers, multiple channels for EMRI formation exist.  The best-studied of these channels is two-body relaxation in quiescent galactic nuclei \citep{hopman05, broggi22, qunbar23}, but even here theoretical EMRI rates are uncertain by at least three orders of magnitude \citep{babak17}.  An alternative EMRI formation channel is migration of stellar-mass BHs through AGN discs \citep{kocsis11, pan21a, pan21b, der23}.  EMRI rates from this channel will depend sensitively on migration physics, and especially on the existence of migration traps, which can choke off the supply of EMRIs to the inner MBH.

In this paper we examine the onset and importance of thermal torques applied to a massive body embedded in an AGN disc. We identify the parameter space where thermal torques are most likely to dominate over the standard gravitational torques, leading to outward migration. Although there is some dependence on the $\alpha$ viscosity and the accretion rate, the mass of the central MBH essentially determines the sign of the total torques: lower-mass MBH discs are prone to outward migration. Conversely, high-mass discs do not exhibit any migration traps and the migration is always inward. The reversal of migration for lower-mass discs leads to the accumulation of BHs much further from the MBH than previously expected.

This paper is organised as follows: In sec. \ref{2} we construct an analytic disc model based on Shakura-Sunyaev $\alpha$-disc theory. In sec. \ref{3} we derive the conditions for which thermal torques change sign and dominate over gravitational torques. In sec. \ref{4} we explore the parameter space and locate the existence and radial location of migration traps and anti-traps (unstable locations where the total torque vanished). In sec. \ref{5} we discuss the implications of these results for the AGN channel for BH mergers, along with other prospects and caveats. Finally, in sec. \ref{6} we summarise our main results.

\section{AGN disc structure} \label{2}

Here we provide a simple prescription for the AGN disc structure, which extends the analytical construction of the \cite{ss73} model to include other effects, such as self-gravity and
low-temperature opacities that have been addressed in more detailed AGN models \citep{sg03,tqm05,gilbaum2022}\footnote{While this paper was under review, a recent preprint appeared \citep{pAGN} which solves both the \cite{sg03} and \cite{tqm05} models, discusses their differences in detail and implements our thermal torque prescription.}.

AGN discs are generally modelled as thin accretion discs. On galactic scales, the discs are cold, gravitationally unstable and fragment into stars, with the effective Toomre $Q_{\rm T}=\Omega^2/2\pi G \rho$, where $\Omega=(GM_\bullet / R^3)^{1/2}$ s the Keplerian frequency\footnote{Throughout this paper, we focus on radii small enough that the overall potential is dominated by that of the massive BH.} and $\rho$ is the density. However, only a fraction of the gas is fragmented and the remaining gaseous component is retained. The transition from the unstable galactic discs to accretion discs is still unclear. Angular momentum transport via global torques, rather than local viscosity, may keep the disc marginally stable and avoid fragmentation.  This mechanism is assumed in the \cite{tqm05} throughout the disc, even at smaller radii, which leads to a relatively cold and underdense disc due to the enhanced viscosity.

The inner structure of AGN discs can be inferred from the optical/UV spectral energy distribution (SED) in the inner regions, and from mass inflow and star formation in the outer regions. \cite{sg03} obtained an accretion disc model that fits the SED observations, while \cite{tqm05} constructed a model regulated by mass inflow and star formation. Recently \cite{gilbaum2022} had shown that accretion feedback from large populations of compact objects can lead to steady-state solutions with $Q_{\rm T} \gg 1$ without active star formation.  Here, we will construct a simple self-regulated model which will be somewhat similar to that of \cite{sg03} {(which is more physically motivated for the inner regions of interest)}, but with simplified analytical expressions for the different parameters. This allows for a broad parameter study of migration torques in discs with different masses, accretion rates, and viscosity laws.


In the following, we use the notation of \cite{ss73} with minor changes, namely that the cylindrical radius $R$ is normalised to the innermost-stable-circular-orbit (ISCO) for a non-rotating BH, $r=R/r_{\rm ISCO}= R/6r_g$, where $r_g=GM_\bullet/c^2$ is the gravitational radius. The mass of the SMBH, $M_\bullet$, is normalised to $10^8M_\odot$, i.e. $m=M_\bullet/10^8M_\odot$, and the accretion rate is normalised to the critical accretion rate, $\dot{m} = \dot{M}/ \dot{M}_{\rm cr}$. Here, $\dot{M}_{\rm cr}=\mathcal{L}_{\rm Edd}/\eta c^2$ is the rate at which the luminosity of the AGN, $L=\eta \dot{M}c^2$, matches the Eddington luminosity $\mathcal{L}_{\rm Edd}=4\pi GM_\bullet c/\kappa$. Here, $G,c$ and $\kappa$ are the gravitational constant, speed of light and opacity, respectively. We take $\eta=0.06$, the accretion efficiency for a non-rotating black hole, but note that it can be larger for a rotating black hole. In our units, $\dot{M}_{\rm cr} = 3 (\eta/0.06)^{-1} m\ \rm M_\odot\ yr^{-1}$.

The inner region (zone I) is radiation-dominated ($P=P_{\rm rad}$), with opacities dominated by electron scattering: $\kappa = \kappa_{\rm es} = 0.2(1+X)\ \rm cm^2\ g^{-1}$, where $X$ is the hydrogen fraction. At larger radii, once the temperature is cool enough, gas pressure will overcome radiation pressure (zone II, $P=P_{\rm gas}$, $\kappa=\kappa_{\rm es}$). 
For even cooler temperatures, the  opacity will become dominated by bound-free and free-free absorption, approximated by Kramers' law: (zone III, $\kappa = \kappa_{\rm K})$, where  $\kappa_{\rm K}=4\times 10^{25} (1+X)Z\rho T^{-7/2} \ \rm cm^2\ g^{-1}$, where $Z$ is the metallicity and the mass density $\rho$ and midplane temperature $T$ are given in cgs units. Throughout the paper, we follow \cite{gilbaum2022}'s parameters for the composition of the gas ($X=0.7381,
    Y=0.2485$ and $Z=0.0134$.

At temperatures in the range $4-8\times 10^3 ~\rm K$, the opacity is dominated by the $H^-$ anion, an opacity source which drops sharply with decreasing temperature, $\kappa_{H^-}=1.1\times10^{-25}Z^{1/2} \rho^{1/2} T^{7.7}  \ \rm cm^2\ g^{-1}$. More realistic disk models using tabulated opacities find that solutions in this temperature range are not realised in practice, and that disk structure jumps discontinuously to a regime of lower temperatures \citep{tqm05, gilbaum2022}.  We therefore keep the solution for zone III until the $H^-$ opacity drops below the molecular opacity $\kappa_{\rm m}=0.1Z\ \rm cm^2\ g^{-1}$, beyond which point $\kappa_{\rm m}$ will dominate (zone IV).

At some point, the density will become low enough that the disc is gravitationally unstable; at all larger radii, we assume that the disc self-regulates to a condition of marginal stability, i.e. $Q_{\rm T}=\Omega^2/2\pi G \rho=1$, (zone V). In this case, the structure is determined from the $Q_{\rm T}=1$ condition. Importantly, the opacity in this zone does not play a role in the structure of this zone
but only in its thermal properties. In this zone, we approximate the total Rosseland mean opacity using a formula from \cite{metzger17}:

\begin{equation}
\kappa = \kappa_{\rm m} + \left( \kappa_{H^-}^{-1} + (\kappa_{\rm es} + \kappa_{\rm K})^{-1} \right)^{-1}, \label{eq1-kappa_m17}
\end{equation}
For the lower mass MBH discs, the transition to $Q_{\rm T}=1$ will occur further away in cooler regions and so the opacity will be molecular. For higher mass MBHs, the transition to $Q_{\rm T}=1$ behaviour can occur at higher temperature opacities. \cite{sg03}'s model has multiple solution for the opacity in the outer zones, while our analytic model has a unique, continuous opacity law. We compare our models with more detailed solution following \cite{gilbaum2022} and find good correspondence. Figure \ref{fig:11} in appendix \ref{appendix} shows that the opacity is much higher due to dust formation for lower mass models, and our analytic prescripion has limited applicability there.

The midplane density in each zone (in cgs units) is given by 
\begin{align}
\rho(r)=\begin{cases}
3.29\times10^{-14}\alpha^{-1}m^{-1}\dot{m'}^{-2}r^{3/2} & \rm{I} (P = P_{\rm rad}, \kappa=\kappa_{\rm es})\\
7.49\times10^{-6}\alpha^{-7/10}m^{-7/10}\dot{m'}^{2/5}r^{-33/20} & \rm{II} (P\to P_{\rm gas})\\
3.5\times10^{-5}\alpha^{-7/10}m^{-7/10}\dot{m'}^{11/20}r^{-15/8} & \rm{III} (\kappa\to \kappa_K)\\
3.96\times10^{-5}\alpha^{-7/10}m^{-7/10}\dot{m'}^{2/5}r^{-33/20} & \rm{IV} (\kappa \to \kappa_m)
\\
4.5\times10^{-2}m^{-2}r^{-3} & \rm{V} (Q_{\rm T}=1),
\end{cases}
\end{align}
where $\dot{m'}=\dot{m}(1-r^{-1/2})$.
Similar power-law scalings are also derived for the other parameters, e.g., the surface density is given by 
\begin{align}
    \Sigma(r)=\begin{cases}
8.57\alpha^{-1}\dot{m'}^{-1}r^{3/2} & {\rm I}\\
3.17\times10^{6}\alpha^{-4/5}m^{1/5}\dot{m'}^{3/5}r^{-3/5} & {\rm II}\\
9\times10^{6}\alpha^{-4/5}m^{1/5}\dot{m'}^{7/10}r^{-3/4} & {\rm III}\\
9.7\times10^{6}\alpha^{-4/5}m^{1/5}\dot{m'}^{3/5}r^{-3/5} & {\rm IV}\\
10^{9}\alpha^{-1/3}m^{-2/3}\dot{m}'^{1/3}r^{-3/2} & {\rm V}
\end{cases}
\end{align}
The aspect ratio is derived from the vertically-averaged relation $\Sigma=2\rho H$, such that

\begin{align}
    H=\begin{cases}
1.3\times10^{14}m\dot{m}' & {\rm I}\\
2.13\times10^{11}\alpha^{-1/10}m^{9/10}\dot{m'}^{1/5}r^{21/20} & {\rm II}\\
1.27\times10^{11}\alpha^{-1/10}m^{9/10}\dot{m'}^{3/20}r^{9/8} & {\rm III}\\
1.22\times10^{11}\alpha^{-1/10}m^{9/10}\dot{m'}^{1/5}r^{21/20} & {\rm IV}\\
1.16\times10^{10}\alpha^{-1/3}m^{4/3}\dot{m}'^{1/3}r^{3/2} & {\rm V}
\end{cases}
\end{align}

The other quantities follow immediately: The sound speed is $c_s=\Omega H$, where $\Omega(r)=2\times 10^{-3}m^{-1}r^{-3/2}\ \rm s^{-1}$ is the Keplerian frequency. The total pressure is given by $P=c_s^2\rho$, 
which also determines the central temperature. The optical depth is given by $\tau = \kappa \Sigma /2$. 

The transitions between the different zones occur at the radii:
\begin{align}
& R_{{\rm I}\to{\rm II}}	 =449.842\alpha^{2/21}m^{2/21}\dot{m'}^{16/21} \label{eq:r12} \\
& R_{{\rm II}\to{\rm III}}	=987.891\dot{m'}^{2/3}\\
& R_{{\rm III}\to{\rm IV}}	=3333.6\alpha^{-0.28}m^{-0.28}\dot{m'}^{0.385}\\
& R_{{\rm I}\to{\rm V}}	=498\alpha^{2/9}m^{-2/9}\dot{m'}^{4/9}\label{eq:r15}
\\
& R_{{\rm II}\to{\rm V}}	=634.4\alpha^{14/27}m^{26/27}\dot{m'}^{-8/27}\label{eq:r25}
\\
& R_{{\rm III}\to{\rm V}}	=580.65\alpha^{28/45}m^{-52/45}\dot{m'}^{-22/45}\label{eq:r35}
\\
& R_{{\rm IV}\to{\rm V}}	=184.7\alpha^{14/27}m^{26/27}\dot{m'}^{-8/27}\label{eq:r45}
\end{align}
\begin{figure*}
    \centering
    \includegraphics[width=0.97\textwidth]{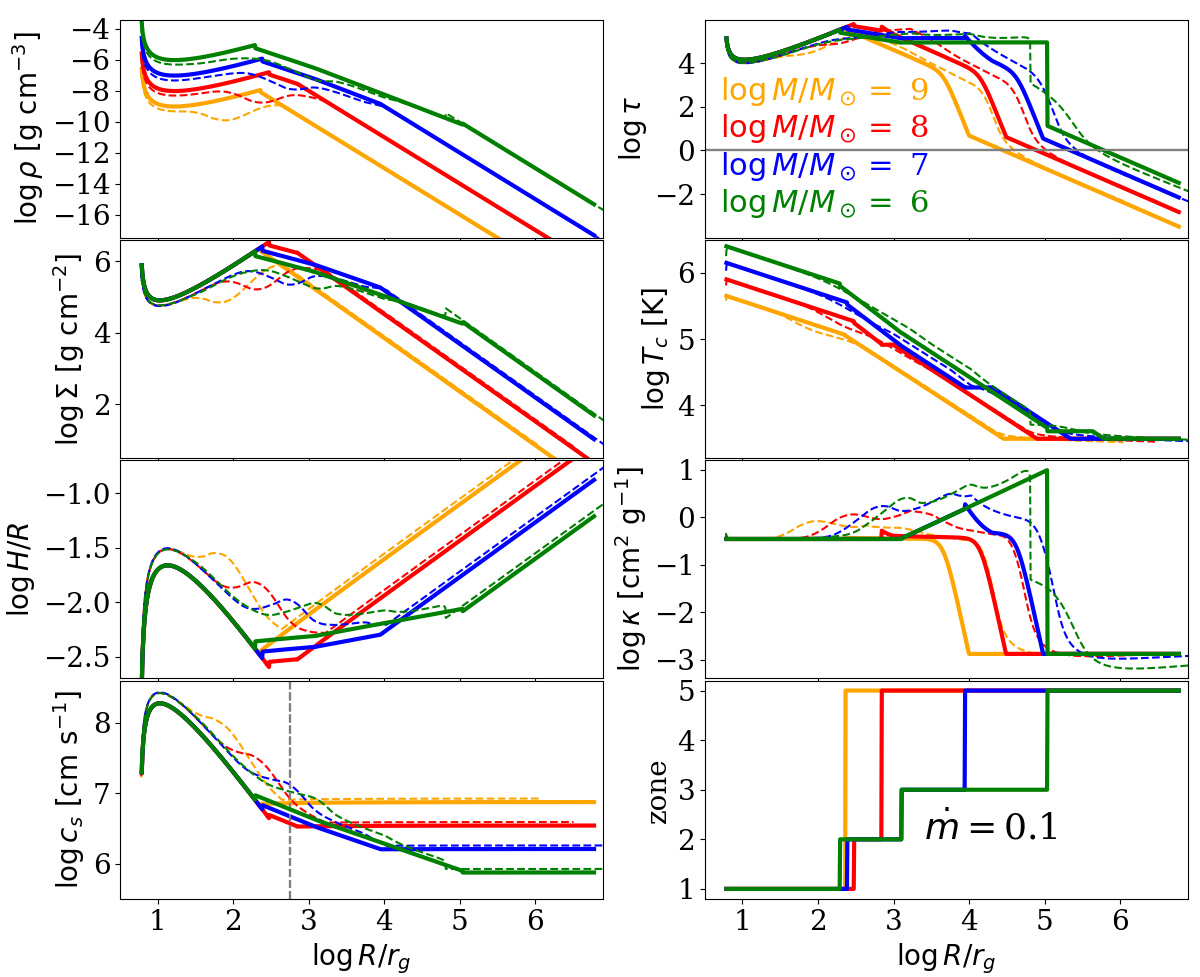}
    \caption{Shakura-Sunyaev type solutions for AGN discs. The different colors represent different SMBH masses as labeled in the legend. Thick solid lines show asymptotic power-law solutions for different ``zones'' (including the $Q_{\rm T}=1$ zone). The thin dashed lines are numerical solutions with tabulated opacities \citep{opal,sem03}. The accretion rate is set to $0.1\dot{m}_{\rm cr}$, and $\alpha=0.01$. The log scales are base 10. {\it Left panels, top to bottom}: volumetric density $\rho$, surface density $\Sigma$, aspect ratio $H/R$, sound speed $c_s$. {\it Right panels, top to bottom}: optical depth $\tau$, central temperature $T_c$, opacity $\kappa$, zone number.  The power-law zones provide a generally good fit to more exact disc structures, with the largest discrepancies coming from bound-free and bound-bound ``opacity bumps.'' }
    \label{fig:1}
\end{figure*}

For a given disc model, only one of the 
transitions to zone V in Eq. (\ref{eq:r15} - \ref{eq:r45}) will occur. Massive discs will transition directly from zone I (Eq. \ref{eq:r15}), while lower mass discs will have progressively more zones \citep{InayoshiHaiman16}.

Figure \ref{fig:1} shows the structure of a disc with dimensionless accretion rate $\dot{m}=0.1$ and dimensionless viscosity parameter $\alpha=0.01$. Each colour represents a different SMBH mass, as labeled in the panel for the 
optical depth $\tau$. The thin lines are numerical solutions that use OPAL opacity tables \citep{opal} at high temperatures and \citet{sem03} tables for low temperature opacities. The larger opacities in the hotter zones are primarily due to iron-line bound-bound and bound-free opacities, which are absent in the power-law $\kappa_{\rm K}$ approximation.

From Figure \ref{fig:1}, we see that the density $\rho$ decreases at fixed $r$ with increasing SMBH mass ($\rho \propto m^{-1}$ for zone I and $\rho \propto m^{-7/10}$ for zones II-III). This causes the transition to the $Q_{\rm T}=1$ zone to occur further away in units of $r_g$, since generally $Q_{\rm T}\propto \Omega^2 /\rho \propto m^{-2}/\rho$. The transition to zone 
V occurs directly from zone II for $M=10^8M_\odot$ (red line), from zone III for $M=10^6, 10^7M_\odot$ (green and lines respectively), and directly from zone I for $M=10^9M_\odot$ (orange line). 

The pressure becomes gas dominated at zone II and remains so until after the disc becomes gravitationally unstable. Thus, in zone V, the central temperature $T_c$ is initially flat. The steep decrease in density makes the pressure radiation dominated again at larger radii, which further reduces $T_c$ as $T_c \propto r^{-3/4}$, Finally, once the disc becomes optically thin, $T_c$ again becomes flat, similar to the \cite{sg03} solution. Only the $M_\bullet=10^9M_\odot$ case has radiation pressure dominance throughout the disc, since it becomes gravitationally unstable before it is able to transition to zone II.

Throughout this paper, we employ the simple power-law model presented here for the AGN disk structure, rather than numerically solving a set of equations with tabulated opacities.  Our motivation for this is three-fold: first, we achieve reasonable agreement with the tabulated opacity solutions (see Figure \ref{fig:1}).  Second, the simple analytic nature of a multi-zone model is computationally cheap and allows for more physical insight.
Third, as we check later on, the migration traps that are our main object of study seem qualitatively robust to choice of disc structure (i.e. power-law models versus more realistic models).

\section{Torques in AGN discs} \label{3}
Having established the structure of an AGN disc for a given a set of initial parameters $\{m, \dot{m}$, and $\alpha$\}, we can now estimate the strength of different migratory torques.  We note that the hydrodynamic torques considered here only apply to embedded objects on prograde orbits (as e.g. Lindblad and corotation resonances will fail to exist for retrograde orbiters). 
 The fraction of all embedded BHs moving on prograde orbits is uncertain but likely greater than $50\%$, as stars and compact objects that form {\it in situ} will inherit their orbital angular momentum from the parent gas disc \citep{stone17}, while alignment of BHs on inclined orbits gives a more even split \citep{nasim23}\footnote{While this paper was under review, \cite{wang24} contested the result of \cite{nasim23} and shown that the vast majority of captured BHs will be prograde.}. Unless stated otherwise, the mass of the migrating stellar-mass BH is $m_{\rm BH} = 10 M_\odot$.

\subsection{Updated type I migration torques}
Consider a massive body, e.g., a stellar mass BH with mass $m_{\rm BH}$ embedded in an accretion disc at cylindrical radius $R$ with surface density $\Sigma(R)$. The gravitational interaction of the perturber with the disc forms two overdense spiral arms. Each arm produces a torque on the body with an approximate magnitude
\begin{equation}
\Gamma_{0} = q^{2}\Sigma R^{4}\Omega^{2} \left( \frac{H}{R} \right)^{-3}, \label{eq: Gamma_0}
\end{equation}
where $q=m_{\rm BH}/M_\bullet$ is the mass ratio between the body and the central SMBH. The two spiral arms compete against each other, and their respective torques cancel to leading order. The cancellation is not complete, however, and the residual net torque is typically (i) weaker by a factor of $H/R$ and (ii) negative in sign, which leads to inward migration. The net torque is \citep[e.g.,][]{paardekooper2010} 
\begin{equation}
    \Gamma_{\rm I} = C_{\rm I}\frac{H}{R} \Gamma_0,
\end{equation}

The prefactor $C_{\rm I}$ depends on the density and temperature gradients, $\nabla_T \equiv - d\ln T_c/d\ln r$ and $\nabla_\Sigma =- d\ln \Sigma / d\ln r$. 
One version of $C_{\rm I}$, widely used in the context of AGN disks with embedded BHs \citep[e.g.][]{bellovary16, gilbaum2022}, is derived in \cite{paardekooper2010}.  However. a more recent revision that is calibrated from 3D hydrodynamical simulations is given in \citet{jm17}, with significantly different astrophysical implications.
The different prefactors we examine are 

\begin{align}
C_{\rm I}=\begin{cases}
-0.85-0.9\nabla_{\Sigma}-\nabla_{T} & \mathtt{p10}\\
C_{{\rm L}}+(0.46-0.96\nabla_{\Sigma}+1.8\nabla_{T})/\gamma & \mathtt{jm17\underline{\ }lin\underline{\ }tot}\\
-1.36-0.54\nabla_{\Sigma}-0.5\nabla_{T} & \mathtt{jm17\underline{\ }lin\underline{\ }iso} \label{eq14-jm17}
\end{cases}
\end{align}
where $\gamma$ is the adiabatic index. $C_{\rm L} = (-2.34+0.1\nabla_{\Sigma}-1.5\nabla_{T})f_{\gamma}$ is given in Eq. in \cite{jm17}. $\chi$ is the thermal diffusivity. For radiative diffusion, the expression is \citep{paardekooper2014}
\begin{align}
\chi & =\frac{16\gamma(\gamma-1)\sigma_{\rm SB} T_c^{4}}{3\kappa\rho_{0}^{2}H^{2}\Omega^{2}}\label{chi-rad},
\end{align}
where $\sigma_{\rm SB}$ is the Stefan-Boltzmann constant and $T_c=T_c(r)$ is the midplane temperature profile.  Here we use the ambient or ``average'' opacity of the AGN disc solution to as the $\kappa$ value used to compute $\chi$, but we note that the relevant opacity could be altered (in most cases, reduced, thus increasing $\chi$) if accretion feedback from the small BH produces a localized hot spot around it.  Such a situation would arise if the number of embedded BHs in the AGN disc is relatively low; alternatively, if the self-regulated discs of \citet{gilbaum2022} are achieved (where BHs continue forming until their accretion feedback efficiently mixes throughout the disc), then such localized hot spots would not exist. The function $f_\gamma$ is given by 
\begin{equation}
    f_\gamma(x) = \frac{(x/2)^{1/2}+1/\gamma}{(x/2)^{1/2}+1} \to \begin{cases}
1 & x\gg 1\\
1/\gamma & x \ll 1
\end{cases}, \label{f_gamma}
\end{equation}
where $x\equiv \chi / H^2\Omega$ is the ratio of the thermal diffusivity $\chi$ to $H^2\Omega$. The $\mathtt{p10}$ indicates Eq. 49 in \cite{paardekooper2010}\footnote{The softening length to scale height ratio is taken equal to 0.4.}, while $\mathtt{jm17\underline{\ }lin\underline{\ }tot}$, and $\mathtt{jm17\underline{\ }lin\underline{\ }iso}
$ indicate Eqs. 67 and 68 in \cite{jm17}, respectively.  The isothermal torque prefactor $\mathtt{jm17\underline{\ }lin\underline{\ }iso}$ is the limit taken by $\mathtt{jm17\underline{\ }lin\underline{\ }tot}$ in a locally isothermal disc, namely one with large thermal diffusivity $\chi$.

\begin{figure}
    \centering
    \includegraphics[width=0.93\linewidth]{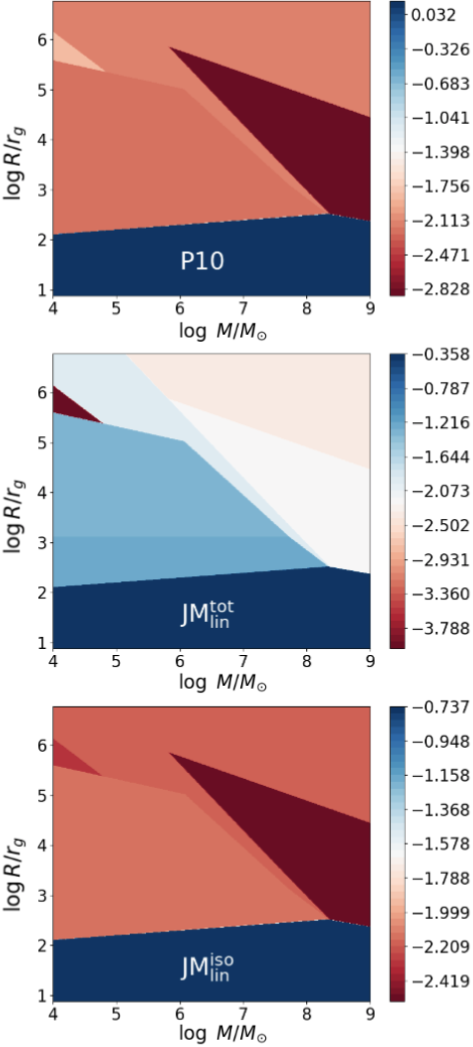}

    \caption{The Type I prefactor $C_{\rm I}$ in Eq. \ref{eq14-jm17} shown with a linear color map for a range of radii $R$ and MBH masses $M_\bullet$, ($\alpha=0.01$ $\dot{m}=0.1$ in all panels). When $C_{\rm I} <0$, embedded objects migrate inwards; when $C_{\rm I} > 0$, migration is outwards.  While the widely used Type I migration formulae of \citet[][\it top panel]{paardekooper2010} robustly produce migration traps at $r \sim 100$, torques are negative-definite in both the general ($\mathtt{jm17\underline{\ }lin\underline{\ }tot}$; {\it middle panel}) torque formulae of \citet{jm17}, and in their isothermal limit ($\mathtt{jm17\underline{\ }lin\underline{\ }iso}$; {\it bottom panel}).}
    \label{fig:2}
\end{figure}

Figure \ref{fig:2} shows the values of $C_{\rm I}$ for different mass ranges and fiducial parameter choices $\alpha=0.01$, $\dot{m}=0.1$. We see that the \texttt{p10} formula indeed yields a positive value in the inner, radiation-dominated zone, naturally producing Type I migration traps. In this zone, $\nabla_\Sigma=-3/2$ (positive slope) and $\nabla_T=3/8$, so $C_{\rm I}=0.17$. The revised formulae of \citet{jm17} have a steeper scaling with the temperature, which outcompetes with the positive density gradient. For both the isothermal, $\mathtt{jm17\underline{\ }lin\underline{\ }iso}$, and general $\mathtt{jm17\underline{\ }lin\underline{\ }tot}$ formulae, the revised torques are always negative, eliminating Type I migration traps. We have confirmed that the type I torque sign remains negative-definite not just for our fiducial power-law disk model but also for a more realistic model using tabulated opacities. \cite{der23} arrived at similar conclusions with isothermal discs and piecewise opacities (see their appendix A) in the context of in-situ inspirals in gravitationally unstable discs.

While the use of updated Type I torque formulae eliminates the migration traps found in past work \citep{bellovary16, secunda19, gilbaum2022} to exist at scales $r \sim 100$, we have yet to consider the additional role of thermal torques.  In the next subsection, we will see that they are capable of re-establishing migration traps, albeit at larger distances.

\subsection{Thermal Torques} \label{2.1}

\begin{figure*}
      \centering
    \includegraphics[width=0.83\textwidth]{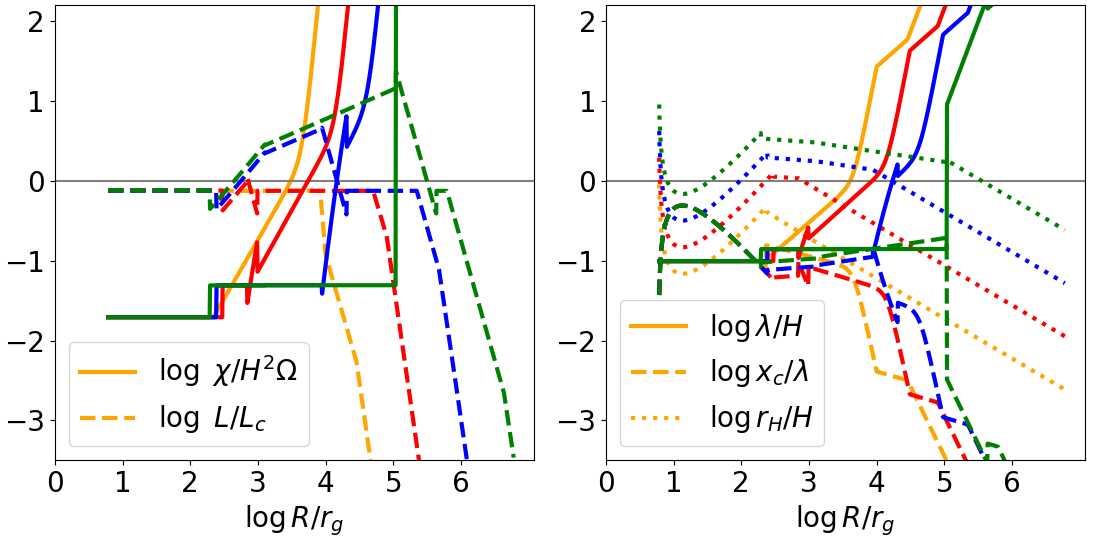}
    \caption{Dimensionless quantities relevant for migration, with colors representing the different MBH masses of Figure \ref{fig:1}. {\it Left panel}: the thermal diffusivity $\chi$ in units of $H^2\Omega$ (solid lines) and the ratio of the stellar BH (Eddington-capped)
    luminosity to the critical luminosity (dashed lines) are plotted against radius $R$. When $L>L_{\rm c}$ ($L<L_{\rm c}$), thermal torques are generally positive (negative).{\it Right panel}: key ratios determining self-consistency of different torque formulae.  Here we show the ratios of the diffusion length $\lambda$ to $H$ (solid), the corotation radius $x_c$ to $\lambda$ (dashed), and the stellar BH Hill radius, $r_H=R(m_{\rm BH}/3M_\bullet)^{1/3}$, to the scale height $H$ (dotted). Dimensionless ratios less than unity (in the {\it right panel}) indicate self-consistency of hydrodynamic torque formulae. }
    \label{fig:3}
\end{figure*}

Aside from gravitational (Type I) torques, ``thermal torques'' 
that originate from heat diffusion near a massive perturber may also affect its orbital evolution. Without the release of heat, two cold and dense lobes are formed which leads to additional inward migration \citep{lega14}. For a body that releases heat back into the ambient gas, the effect is the opposite, and two hot, under-dense lobes are formed \citep{b-l15}.  \citet{masset17} studied the two effects analytically and combined their contributions in terms of the {\it total heating torque}. While the one-sided thermal torque is also comparable to $\Gamma_0$, the total thermal  torque is found to be (Eq. 145 in \citealp{masset17})
\begin{equation}
    \Gamma^{\rm tot}_{\rm thermal} = 1.61 \frac{\gamma-1}{\gamma}\frac{x_c}{\lambda}\left(\frac{L}{L_c} -1\right) \Gamma_0 \label{eq15-tot_th_torque},
\end{equation}
where $x_c$ is the body's corotation radius, $\lambda$ is the typical size of the lobes, and $L$ is the emerging luminosity generated by the massive body. The quantity $L_c$ is the critical luminosity (Eq. 132 in \citealp{masset17}):
\begin{equation}
    L_c = \frac{4\pi Gm_{\rm BH}\rho} \gamma{\chi}.\label{L_c}
\end{equation}
If the emerging luminosity from the massive body equals the critical luminosity, $L=L_c$, the hot and cold torques exactly balance each other and cancel out. For a non-luminous body, with $L=0$, only the cold torque is present. 

If the emerging luminosity is a fraction $\epsilon_{\rm Edd}$ of the Eddington luminosity, $L=\epsilon_{\rm Edd} L_{\rm Edd} =\epsilon_{\rm Edd} \times 4\pi Gm_{\rm BH}c/\kappa$, then the condition for a positive thermal torque is $L>L_c$ or 

\begin{equation}
    \frac{\epsilon_{\rm Edd} \gamma c}{\kappa \rho \chi} >1 \implies  \chi \sim \lambda_c^2 \Omega < \epsilon_{\rm Edd} \gamma c \ell. \label{eq: sign-change}
\end{equation}
 Here $\kappa$ is the opacity, $c$ is the speed of light and $\ell \equiv 1/\kappa \rho$ is the photon mean free path. In order to estimate where this condition is met, we need to specify the structure of the accretion disc and the origin of the thermal diffusivity.

 The prefactor $\epsilon_{\rm Edd}$ is set to be the minimum value between $1$ and $L_{\rm rBHL}/ L_{\rm Edd}$, where $L_{\rm rBHL}$ is the reduced Bondi-Hoyle-Littleton accretion luminosity, $L_{\rm rBLH} = \eta c^2 \times \pi S_{\rm rBHL} c_s\rho$. The effective reduced cross-section, $S_{\rm rBHL}$ depends on the relations between the Hill radius, scale height and the classical bondi radius and is given in Eq. 20 of \cite{gilbaum2022}. In practice, $\epsilon_{\rm Edd} \ll 1$ only for the outer regions of high-mass discs. These regions tend to be optically thin, where thermal torques are significantly quenched.

\subsubsection{Length scales, dimensionless numbers and assumptions} \label{assumptions}
In addition to the scale height $H$ of the disc, the length of the lobes is estimated by comparing the diffusion timescale $\lambda^2/\chi$ to the Keplerian shear timescale $1/\Omega$, which leads to

\begin{equation}
    \lambda \equiv \sqrt{\frac{2\chi} {3\gamma \Omega} }. \label{eq-18-lambda}
\end{equation}
The corotation radius of body $m_{\rm BH}$ is the location where the gas rotates at the same speed as the body. It is non-zero because the radial pressure gradient slightly reduces the gas velocity, and its value is
\begin{equation}
    x_c \equiv \frac{\nabla_P c_s^2}{3\gamma R\Omega^2} = \frac{\nabla_P}{3\gamma} \frac{H^2}{R} ,\label{eq-19-xc}
\end{equation}
which is much smaller than the scale height. In addition, the Hill radius, $r_H\equiv r (M_\bullet/3m_{\rm BH})^{1/3}=r(1/3q)^{1/3}$ measures the importance of the gravitational pull of the stellar mass BH.

The linear theory used in analytic derivations of thermal \citep{masset17} torques requires the expansion of the gravitational potential in powers of $x_c/\lambda$, thus $x_c\ll \lambda$ is a stringent requirement for their validity. A less stringent self-consistency requirement is that $\lambda \ll H$, a condition that validates the shearing-sheet approximation. However, even if $\lambda$ is comparable to $H$, the hydrodynamical simulations of \cite{hankla2020} find good agreement with the linear theory. 

The left panel of Figure \ref{fig:3} shows the thermal diffusivity $\chi$ in units of $H^2\Omega$. We see that outside of the $Q_{\rm T}=1$ zone, $\chi$ is a constant, small fraction of $H^2 \Omega$, on the order of $\alpha$. The "jumps" are from discontinuous changes of $\gamma$ whenever they are applicable (i.e., $\gamma=5/3$ when $P=P_{\rm gas}$ and $\gamma=4/3$ when $P=P_{\rm rad}$).  In this case, the disc luminosity is connected to its accretion, so (Eq. 3 and 4 in \citealp{sg03}):
\begin{equation}
    \sigma_{\rm SB} T_c^{4}=\frac{3\sigma_{\rm SB}\tau}{8}T_{{\rm eff}}^{4}=\frac{3\tau}{8\pi}\frac{3}{8}\dot{M}  \left( 1- r^{-1/2} \right) \Omega^{2}, \label{T profile}
\end{equation}
where $T_{\rm eff}$ is the effective temperature. The accretion rate is governed by viscous transport, i.e. (Eq. 6 in \citealp{sg03}) $\dot{M}  \left( 1- r^{-1/2} \right)  = 3\pi \nu \Sigma$. The viscosity $\nu=\alpha c_s H = \alpha H^2 \Omega$ is parametrised by the Shakura-Sunyaev $\alpha$ prescription. 
Plugging in the expression for $\sigma_{\rm SB} T_c^4$ and for the accretion rate in terms of the viscosity into the expression for radiative diffusion (Eq. \ref{chi-rad}), the thermal diffusivity is given by
\begin{equation}
    \chi = \frac{9\gamma(\gamma-1)}{2} \alpha H^{2}\Omega \label{chi-fin}.
\end{equation}
The ratio of viscous to thermal diffusion is defined as the Prandtl number ${\rm Pr}\equiv \nu/\chi$, which is of order unity for the inner $Q_{\rm T}\gg1$ zones (${\rm Pr}=1/5$ for $\gamma=5/3$, and ${\rm Pr}=1/2$ for $\gamma=4/3$). The optically thin and $Q_{\rm T}=1$ zone are characterised by small Prandtl numbers. 

\begin{figure*}
      \centering
    \includegraphics[width=0.9\textwidth]{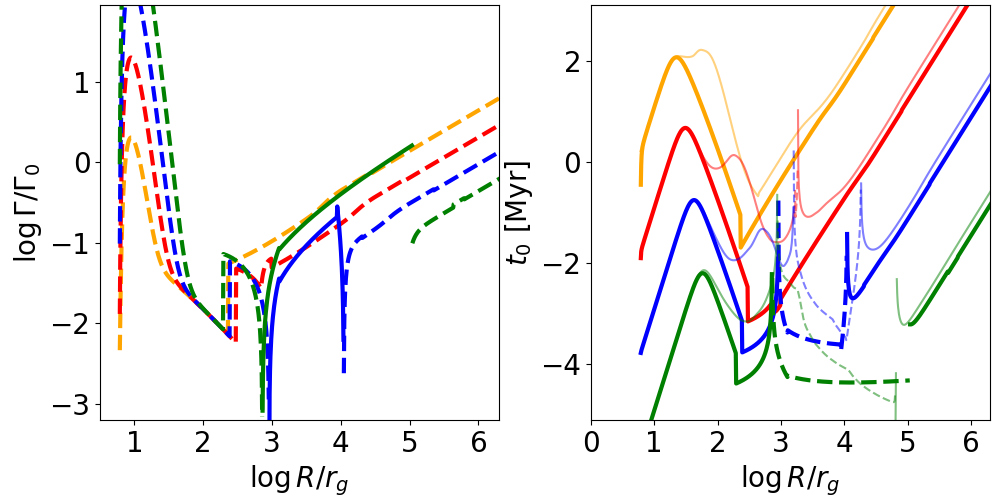}
    \caption{{\it Left panel}: the total torque $\Gamma$ normalised to $\Gamma_0$. Dashed lines represent regions of negative torque and inward migration, while solid lines show regions of positive torque and outward migration. The colours label different MBH masses in the same way as in Figure 1. {\it Right panel}: migration times $t_0$ of BHs embedded in the discs. Solid lines are for inward migration, while dashed lines are for outward migration. Thin transparent lines indicate the solution with tabulated opacity, while thick lines indicate the power-law solutions. The power-law disc solutions show generally good agreement with the more precise ones regarding torque sign and the existence or not of migration traps.  The one qualitative disagreement visible here is for the $M_\bullet = 10^8 M_\odot$ (red) case, which has no traps in the power-law disk model and one trap near $R \approx 2000 r_{\rm g}$ in the more realistic modeling.  This discrepancy is an edge case discussed below (e.g. Figure \ref{fig:6}). }  
    
    \label{fig:4}
\end{figure*}

The right panel of Figure \ref{fig:3} shows the ratio of $\lambda/H$ (solid), $x_c/\lambda$ (dashed), and $r_{\rm H}/H$  (dotted). We see that the assumptions for the linear expansion of thermal torques $(x_c < \lambda)$ are generally valid beyond $r\gtrsim 20 r_g$. The assumption of negligible shear $(\lambda<H)$ is valid for $r\lesssim 10^4 r_g$ for $M_\bullet=10^8$ with an approximate dependence of $m^{-1/2}$. This roughly occurs $0.5-1$ dex after the transition to the $Q_{\rm T}=1$ zone. The latter regimes had not been numerically explored and need to be verified. We expect the thermal torques to be smaller than the Type I torques here since the degree of asymmetry for Type I torques, $x_c/H$, will overtake $x_c/\lambda$, the degree of asymmetry for thermal torques. Finally, $r_{\rm H}$ is always smaller than $H$, (at least for a 10 $M_\odot$ BH), so self-gravity can be safely neglected.

\begin{figure*}
      \centering
    \includegraphics[width=0.87\textwidth]{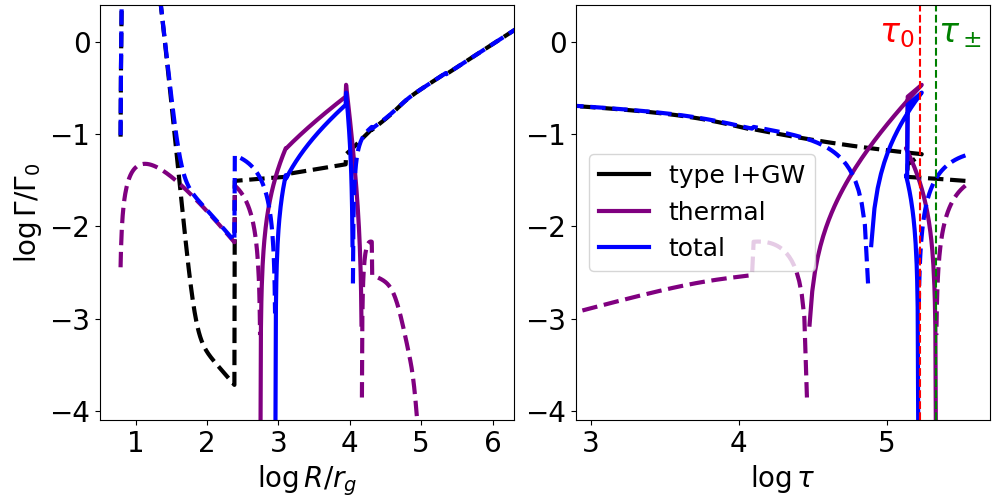}
    \caption{
    Individual normalised torques for the $M_\bullet = 10^7 M_\odot$ case. Black lines show combined type I torques (with the $\mathtt{JM17\underline\ lin\underline\ tot}$ prescription) plus GW torques. Purple lines are the thermal torques, while blue lines are the total torques. The linestyle is the same as in Figure \ref{fig:4} (dashed for inward migration, solid for outwards). {\it Left panel}: torques plotted in $R$-space. {\it Right panel}: torques in $\tau$-space. The dashed vertical lines indicate $\tau_\pm$ in Eq. \ref{taupm} (green) and $\tau_0$ in Eq. \ref{tau-2} (red), locations where migration traps are likely to occur.
    }
    \label{fig:5}
\end{figure*}

\subsection{The total torque}
We are now ready to sum the gravitational and thermal torques into an expression for the total torques and to explore the total torque sign and magnitude. For the thermal torque, we assume that the main driver of thermal perturbations is radiative diffusion. In optically thin regions, this assumption is invalid, since even if radiation remains the main heat transport mechanism, it will not follow the diffusion approximation\footnote{{Note, however, that this conclusion could change with more realistic, frequency-dependent opacities.  For example, cold regions of the AGN disc that are optically thin to their own thermal radiation (in the Rosseland mean opacity sense) may be optically thick to the ionizing X-rays and UV produced by mini-disks around accreting BHs \citep{gilbaum2022}.}} (used in Eq. \ref{chi-rad}). However, the validity of the thermal torque formulae we employ probably requires an even stricter condition on optical depth, namely that at least one scattering/absorption occurs as a photon travels the characteristic distance between the perturber and the thermal lobes, $\lambda$ (i.e. $\ell < \lambda$, or $\tau > H/\lambda$).
From Figure \ref{fig:3}, $H/\lambda$ is largest in the inner zones, where Eq. \ref{T profile} is valid. For this case, $\lambda \sim \sqrt{\alpha} H$, and our condition becomes $\tau > \alpha^{-1/2}$. For practical purposes, in what follows we multiply the thermal torque by $g(\alpha,\tau) = 1- \exp(-\sqrt{\alpha}\tau)$, which ensures that the thermal torque exponentially decays to zero for $\tau \ll \alpha^{-1/2}$, though we note that this is only an approximate treatment of an uncertain regime.

The left panel of Figure \ref{fig:4} shows the total torques obtained for our fiducial (power-law) disc models. We see that the high-mass cases ($M_\bullet = 10^8$ and $10^9 M_\odot$) have negative-definite torques and no migration traps at all. In all cases, there is always inward migration in the innermost zone and in the transition to the next zone (zone II for $M_\bullet \le10^8 M_\odot$ or zone V for $M_\bullet=10^9M_\odot$), which occurs when the torque magnitudes jump around $R=$ few  $\times 100r_g$.  The total torque changes sign first for the low mass cases around $\sim 800 r_g$. The total torque remains positive up to $\sim 10^4\ r_g$ for $M_\bullet = 10^7M_\odot$ (blue lines) and $\sim 10^5\ r_g$ for $M_\bullet = 10^6M_\odot$ (green lines). The torque spike in the latter case is due to the sudden drop in the opacity, as $\Gamma \propto L/L_c \propto \kappa^{-1}$. 

The sharp increase of negative torque for smaller values of $r$ is due to the torque from GW emission \citep{Peters1964} which eventually leads to an extreme mass ratio inspiral,
\begin{equation}
    \frac{\Gamma_{\rm GW}}{\Gamma_0} = \frac{32}{5}\frac{c^{3}}{c_{s}^{3}}\frac{H^{6}}{R^{6}}\left(\frac{r_{g}}{R}\right)^{4}\frac{M}{\Sigma R^{2}}
\end{equation}
Due to the extremely steep dependence on $r$ ($\Gamma_{\rm GW}/\Gamma_0 \propto r^{-12}$), the GW torque is important on short distance scales ($r\lesssim 100$), while any migration traps are located much further away. We include the GW torque in the total gravitational torque shown in Figure \ref{fig:4}, $\Gamma_{\rm tot} = \Gamma_{\rm I}+\Gamma_{\rm thermal}+\Gamma_{\rm GW}$. 

The right panel of Figure \ref{fig:4} shows the migration timescale
\begin{equation}
    t_{\rm mig} = \frac{m_{\rm BH}\sqrt{GMR}}{2\Gamma} = \frac{1}{2\Omega} \left(\frac{H}{R}\right)^3 \frac{M}{\Sigma R^2} \frac{M}{m_{\rm BH}} \left( \frac{\Gamma}{\Gamma_0}\right)^{-1},
\end{equation}
for both the fiducial (power-law) disc models and more accurate ones produced with tabulated opacities. The correspondence is excellent in the outer zones where the disc solutions converge and in the innermost regions where GW torques dominate. The timescale $t_{\rm mig}$ is longer for the tabulated opacity solutions in intermediate regions, due to the reduced surface densities there. 
Deviations from power-law behavior at intermediate radii come primarily from non-Kramers bound-free and bound-bound opacity structures, such as the ``iron opacity bump'' at $T_{\rm c} \approx 10^5~{\rm K}$.  Overall, migration timescales near traps are relatively fast.  While AGN lifetimes $t_{\rm AGN}$ are highly uncertain, they are bounded from above ($t_{\rm AGN} \lesssim 10^8 ~{\rm yr}$) by the quasar luminosity function \citep{HopkinsHernquist09, gilbaum2022}.  Whether this total cosmic lifetime is subdivided into many smaller AGN episodes is unclear, but in at least some systems this is not the case \citep{Martini04}, and there is some evidence that typical episodic lifetimes span a broad range of $\sim 1-30$ Myr \citep{Worseck+21, Khrykin+21}.  Typical values of $t_{\rm mig}$ only exceed these episodic lifetimes well beyond the location of migration traps, especially for 
lower mass discs. 

We conclude that even though the novel population of ``thermal'' migration traps we have found sits significantly further from the SMBH than did the Type I migration traps in past literature \citep{bellovary16, secunda19, gilbaum2022}, they will still be capable of absorbing a large flux of inward-migrating BHs during an AGN episode.

In order to gain more insight, we plot in detail each of the type I, thermal, and total torques for the $M_\bullet=10^7M_\odot$ model in Figure \ref{fig:5}.
We see that when  only Type I torques are present, unlike the  $\mathtt{p10}$ Type I torque,  there are no migration traps (type I torque is always negative, see also Figure \ref{fig:2}).  We get similar results  using   $\mathtt{jm17\underline{\ }lin\underline{\ }iso}$. Regardless of the exact Type I torque prescription, once the thermal torque is introduced, it dominates between $R \approx 800 - 10^4r_g$. The migration trap thus moves to $\sim 10^4r_g$.

\subsection{Critical optical depth} \label{3.4}
Figure \ref{fig:5} plots $\Gamma_{\rm tot}$ not just against $r$ but also against optical depth $\tau$. We show this because the condition for positive (thermal) torque (Eq. \ref{eq: sign-change}) can be expressed as a requirement that $\tau$ is {\bf smaller} than a value
\begin{equation}
    \tau_{\pm} = \frac{\gamma c H}{\chi} 
\end{equation}
If the transition occurs in zones I-IV, Eq. \ref{chi-fin} is valid and we can estimate 
\begin{equation}
    \tau_{\pm}=\frac{2c}{9(\gamma-1)\alpha c_{s}} \approx 10^5\alpha_{0.01}^{-1}c_{s7}^{-1} \label{taupm}
\end{equation}
where the sound speed is normalised via $c_{s7}=c_s/10^7{\ \rm cm\ s^{-1}}$ and the viscosity by $\alpha_{0.01}=\alpha / 0.01$. The expected transition in Eq. \ref{taupm} is seen on the right panel of Figure \ref{fig:5}. 

\begin{figure*}
      \centering
    \includegraphics[width=0.42\textwidth]{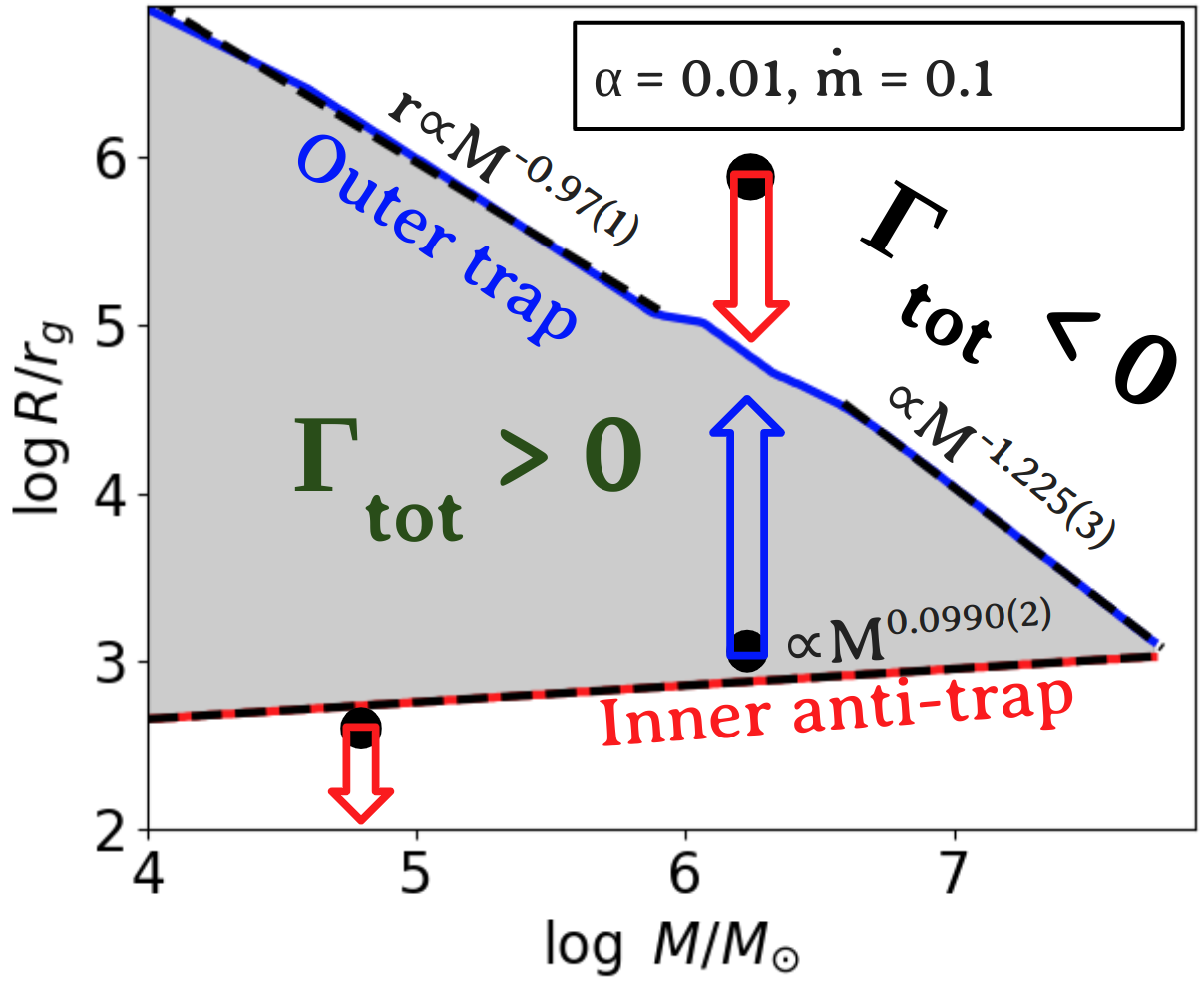}
    \includegraphics[width=0.57\textwidth]{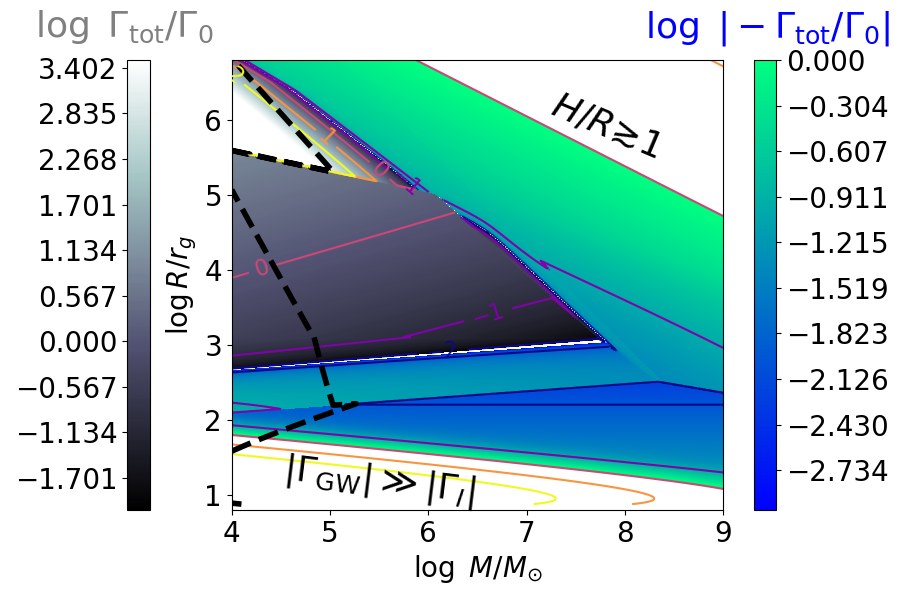}

    \caption{Location of the migration traps versus the SMBH mass. The accretion rate and viscosity are set to fiducial values of $\dot{m}=0.1$, $\alpha=0.01$. {\it Left panel}: The red line is the inner anti-trap, while the blue line is the true outer trap. The grey area is the region where a positive net torque applies, leading to outward migration. The rest of the parameter space has negative net torque and inward migration. The red and blue arrows showcase the direction of migration for massive objects in the AGN disc. The linear-regression slopes are $0.0990(2)$ for the inner anti-trap, and $-0.97(1)$ and $-1.225(3)$ for the low-mass and high-mass outer trap, respectively. {\it Right panel}: The total torque magnitude shown as a color map. The greyscale represents positive torques (corresponding to the grey area in the left panel). The contours shows lines of constant $\Gamma_{\rm tot}/\Gamma_0$. The green-blue colour-scale represents negative torque. The black dashed line indicates the transition to type II migration, which occurs to its left.  The upper right corner does not have a self-consistent disc solution since $H/R\sim 1$ there. For the smallest separations GW torques dominate. }
    \label{fig:6}
\end{figure*}

We can also estimate when the thermal torque overcomes the type I torque. The torque ratio (we consider distances large enough from the MBH that GW torques are neglected) is
\begin{equation}
    \frac{\Gamma_{\rm thermal}}{\Gamma_{\rm I}} = 1.61\frac{\gamma-1}{\gamma}\frac{x_c}{\lambda}\left(\frac{L}{L_c}-1\right)\frac{R}{C_{\rm I} H} \label{G_ratio}.
\end{equation}
Plugging in the expressions for $\lambda$, $x_c$ $L/L_c$ and $\chi$ in zone II using Eqs. \ref{eq: sign-change} \ref{eq-18-lambda}, \ref{eq-19-xc}, and \ref{chi-fin} and setting Eq. \ref{G_ratio} equal to unity, we find that the critical optical depth $\tau_0$ separating thermal torque from Type I torque domination is
\begin{align}
1 &= 0.3\frac{\sqrt{\gamma-1}}{|C_{\rm I}|\gamma^{2}}\frac{\nabla_{P}}{\sqrt{\alpha}}\left(\frac{\tau_{{\rm \pm}}}{\tau_{0}}-1\right) \equiv \mathcal{A} \left(\frac{\tau_{{\rm \pm}}}{\tau_{0}}-1\right). \label{tau_0-1}
\end{align}
Note that $\tau_0 < \tau_\pm$ is required for a valid solution. For our case, the transition is in zone II, with numerical values $\gamma=5/3$, $|C_{\rm I}|=1.275$, $\nabla_{P}=51/20$, and $\alpha=0.01$, and we $\mathcal{A}=1.77$.

The solution to Eq. \ref{tau_0-1} is then
\begin{equation}
    \tau_{0}=\tau_{\pm}(r)\frac{\mathcal{A}}{1+\mathcal{A}} \label{tau-2} 
\end{equation}
The solution depends on $\tau_\pm(r)$, which is achieved at a different location. It scales as $\tau_\pm \propto c_s^{-1} \propto r^{9/20}$. We see that the distance in $r$ space between the two sign changes is roughly $\approx 0.2$ dex, so we multiply Eq. \ref{tau-2} by $10^{0.09}\approx 1.23$. We finally get that for our model
\begin{equation}
    \tau_0 \approx 0.79\tau_\pm, \label{tau_est}
\end{equation}
Which is plotted as the dashed vertical red line in the left panel of Figure \ref{fig:5}. 

\subsubsection{Comparison to \protect\cite{hankla2020} }

Eq. \ref{tau_0-1} and \ref{tau-2} were derived for general $\mathcal{A}$. The two values of $\tau_\pm$ and $\tau_0$ have a similar order of magnitude (as in our case) when $\mathcal{A}$ is of order unity, which occurs when $L/L_c$ is not too large. Alternatively, if $L/L_c\gg 1$, Eq. \ref{tau-2} becomes
\begin{equation}
    \tau_{0}=\tau_\pm(r)\mathcal{A}. \label{tau-3} 
\end{equation}
In this case, $\mathcal{A}\ll 1$, or $\tau_0 \ll \tau_\pm$, which means that the thermal torque is positive for a vast range of parameter space before it starts to dominate (the luminosity ratio has to grow by orders of magnitude). In practice, this situation is rarely achieved, since $\mathcal{A} \sim \mathcal{O}(10^{-1}) \times \alpha^{-0.5}$, so it is expected to be $\gtrsim 1$, unless the viscosity is very high.

However, this was the primary case considered in \cite{hankla2020}. In order to compare to their work, we use Eq. \ref{tau-3}. In this case, the scalings with $\alpha$ and $c_s$ are the same as in \cite{hankla2020}, namely $\tau_0 \le 7\times 10^{-4} \alpha^{-3/2}c_s^{-1}=2.1\times 10^3 c_{s,7}^{-1}\alpha_{0.01}^{-3/2}$, where we assume a fiducial value of $H/R=10^{-2}$ to convert from the sound speed to the Keplerian velocity. 

Even when correcting with Eq. \ref{tau-2}, our value is larger by two orders of magnitude. While some order-unity differences may be attributed to variation in $C_{\rm I}$ and a slightly different $\nabla_P$ ($=51/20$ in our case, versus $3$ in \citealp{hankla2020}), the main reason for the discrepancy is that \cite{hankla2020} use $\Gamma_0$ for the gravitational torque, where we use the total Type I net torque $-C_{\rm I}\Gamma_0 (H/R)$. The value of $\Gamma_0$ corresponds to the one-sided torque that is generated from a single arm of a spiral density wave, while the net torque is $\propto (H/R)\Gamma_0$, since the degree of asymmetry in the two spiral arms scales as $H/R$. The torque $\Gamma_{\rm thermal}$ in Eq. \ref{eq15-tot_th_torque} is the {\it total net torque}, and the degree of asymmetry scales as $x_p/\lambda$ (see Fig 2 of \citealp{masset17} for a graphical representation), which needs to be compared with the total net type I torque. The extra $H/R$ term greatly reduces the net gravitational torque and explains the discrepancy.

\section{Parameter space exploration} \label{4}

\begin{figure*}
      \centering
    \includegraphics[width=0.99\textwidth]
     {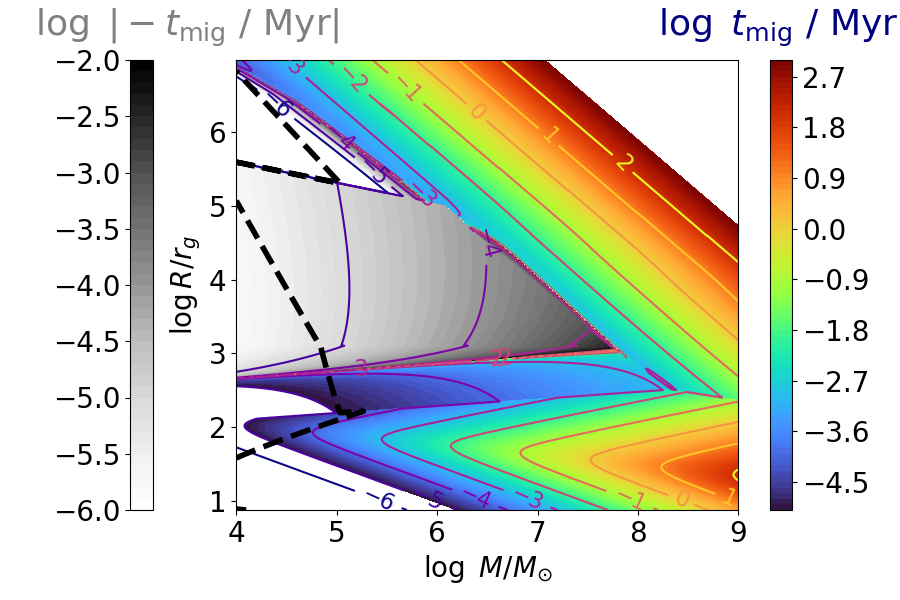}
    
    \caption{ Migration times in $\{M, R\}$ parameter space for a $10 M_\odot$ BH in an AGN disc, with $\dot{m}=0.1$ and $\alpha = 0.01$. The greyscale shows  outward migration (indicated) by negative time, while the rainbow colour-scale is inward migration. Contours of constant $\log_{10} t_{\rm mig}$ in Myr are also indicated  with different colours. The dashed black line is the same as in Figure \ref{fig:6}.  Outwards migration (from positive thermal torques) is generally very rapid in regions where it occurs.}
    \label{fig:7}
\end{figure*}

\begin{figure*}
      \centering
    \includegraphics[width=0.5\textwidth]{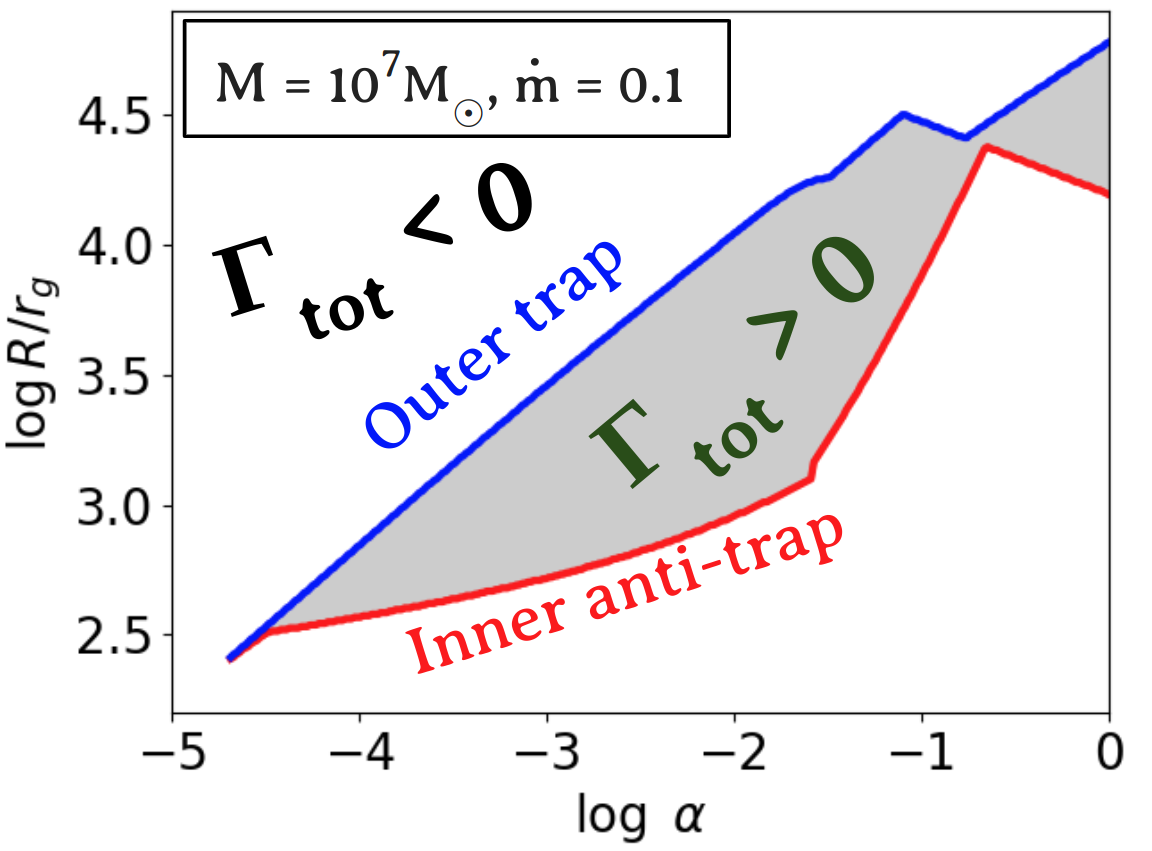}
    \includegraphics[width=0.49\textwidth]{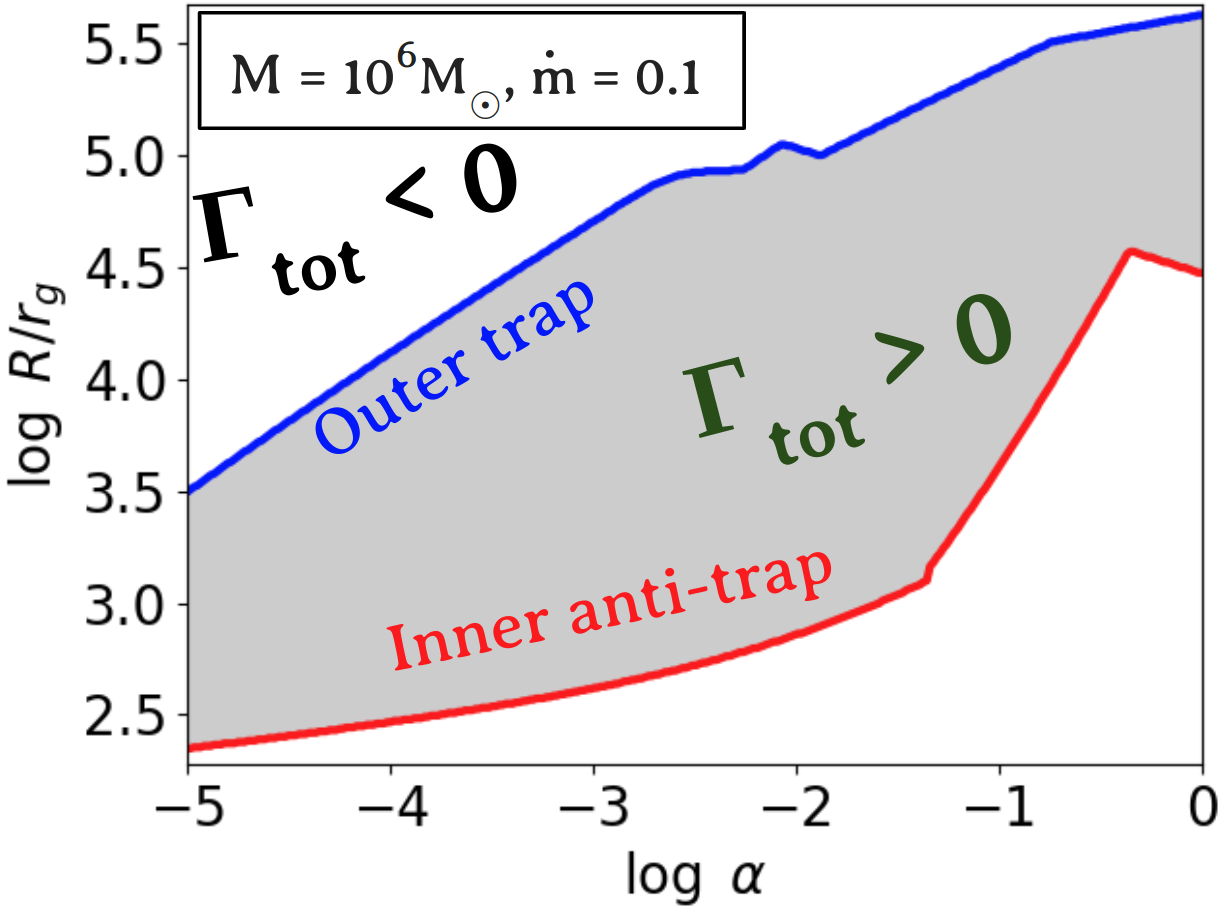}
    \caption{Zones of inward and outward migration for different viscosities $\alpha$. The grey areas indicate locations of positive torques, in the same style as Figure \ref{fig:6}. {\it Left panel}: $M_\bullet=10^7M_\odot$.  {\it Right panel}: $M_\bullet=10^6M_\odot$. The accretion rate in both cases is relatively large, $\dot{m}=0.1$.  Both the outer (true) migration trap and the inner anti-trap generally move outwards with increasing $\alpha$.}
    \label{fig:8}
\end{figure*}

After constructing an AGN disc model with 
initial parameters for $m,\dot{m}$ and $\alpha$, in sec. \ref{2} and calculating the total net torques (both gravitational and thermal) on a massive object in an AGN disc in sec. 3, we were able to provide some analytical constraints in terms of the optical depth on the emergence of torque reversal. We are now ready to explore the different parameters space by varying the disc parameters. Our primary goal will be quantifying the existence and location of points where the migration torque $\Gamma_{\rm tot}=0$, which can produce either a standard migration trap (if $\Gamma_{\rm tot}<0$ outside and $>0$ inside) or an ``anti-trap'' (if the opposite is true).  In general, we find two categories of disc solutions: (i) solutions completely lacking in both migration traps and anti-traps, and (ii) solutions where an inner anti-trap occurs at a radius smaller than an outer migration trap.  Disc solutions of type (ii) are more common, but the relative position of the trap and anti-trap can change dramatically as $M_\bullet$, $\alpha$, and $\dot{m}$ vary.

\subsection{Varying MBH mass}

The left panel of figure \ref{6} shows the radial locations where the torque changes sign as MBH mass varies (for fiducial values $\dot{m}=0.1$, $\alpha=0.01$). The location of the inner anti-trap, $r_1(m)$, is almost constant and scales as $r_1\propto m^{0.099}$.  It is generally located below $r=10^3$, so objects that are formed or captured at $R<10^3 r_g$ will migrate inward to the MBH without obstruction. The true migration trap, located at a radius $r_2(m)$, scales more steeply with $M_\bullet$, either $\propto m^{-1.225(3)}$ for the more massive MBHs, or $\propto m^{-0.97(1)}$ for the less massive ones. 

The power law of the inner anti-trap can be estimated as follows: from the conditions for $\tau_0 \propto c_s^{-1}$ in Eq. \ref{tau-2}, the anti-trap occurs in zone II, where the opacity is constant and $\tau \propto \Sigma \propto m^{1/5}r^{-3/5}$, which needs to be compared with $c_s^{-1}\propto m^{1/10}r^{9/20}$. Comparing the power laws leads to $r\propto m^{2/21}$, which also relates to the scaling of the transition radius, $R_{\rm I\to II}$, in Eq. \ref{eq:r12}.  The location of the outer migration trap, $r_2(m)$, can occur in multiple disc zones, explaining the multiple power laws seen in Figure \ref{fig:6}.  

The grey zone in Figure \ref{fig:6} indicates the region where massive embedded objects migrate outward. It can be understood as part of the ``drainage basin'' or watershed for the outer migration trap at $r_2$.  We see that lower-mass MBHs have more distant migration traps.  For example,
 the $M_\bullet=10^6M_\odot$ trap is located at $r_2 \approx 10^5 r_g$. This means that in the lowest-mass AGNs, the migration proceeds outwards to the edge of our modelled radii.  Notably, the grey region disappears for $M_\bullet \gtrsim 10^8 M_\odot$; in other words, there are no longer any regions of positive-definite torque and all migration is inwards.  At these large MBH masses, low values of $L/L_{\rm c}$ (see Figure \ref{fig:3}) keep the thermal torque negative-definite, preventing the emergence of traps (or anti-traps).

The right panel of Figure \ref{fig:6} shows the total torque map in $R-M$ space. The area of positive torque is again coloured using grayscale, while the coloured regions show the area of negative torque. For small separations, $R\lesssim 20 r_g$, the (negative) GW torques tend to dominate. In the upper right corner, the aspect ratio exceeds unity and the disc becomes a torus, where many of our disc modelling and migration assumptions become untrustworthy. We see that the positive torques tend to be stronger than the negative torques. The black dashed line indicated where the validity of our Type I migration prescriptions break down due to the stellar mass BH opening a gap in the disc and entering the Type II regime (see Sec. \ref{5}). 
 In the upper left corner, the thermal torques are extremely strong, and the linear perturbative analysis may also break down, which is examined in more detail in Figure \ref{fig:7}.

Figure \ref{fig:7} recasts the previous results in terms of migration timescales $t_{\rm mig}$ rather than torques $\Gamma_{\rm tot}$.  Again, the greyscale codes for outward migration, while the rainbow colour bar is for inward migration. We see that outward migration is usually faster and all outward migration occurs on times less than $1$ Myr. The inward migration is also generally fast, so that a BH may migrate within the AGN lifetime of $10^8\ \rm yr$ for $R\lesssim 10^5 (M_\bullet /10^8M_\odot) r_g$. 

We caution that when the timescale for migration becomes comparable to the orbital timescale, the linear migration theories we employ are not valid anymore and our predictions may become untrustworthy (e.g., orbits may become highly eccentric). This predominantly occurs in the upper left "triangle" of Figure \ref{fig:7} where thermal torques are extremely efficient. Moreover, migrating stellar-mass BHs in the $M_\bullet \sim 10^4 M_\odot$ discs tend to form a gap (left of black dashed line) where another assumption on Type I migration becomes invalid.

\subsection{Varying viscosity}

Figure \ref{fig:8} shows the locations of traps/anti-traps as functions of viscosity for two different cases ($M_\bullet=10^6 M_\odot$ and $M_\bullet=10^7 M_\odot$). We see that reducing (increasing) the viscosity moves both the true migration traps and the anti-traps to smaller (larger) values of $R$. 
The maximum separation between the trap and the anti-trap is achieved for intermediate values of viscosity ($\alpha \sim 10^{-2}$, while smaller and larger viscosities shrink this separation.  Very low viscosities (below $\alpha<10^{-4.5}$) lack migration traps altogether, although these are unlikely to be realistic in actual AGN disks.

The two panels of Figure \ref{fig:8} contrast the trap locations between a $M_\bullet=10^7 M_\odot$ and a  $M_\bullet=10^6M_\odot$ case. The zone for positive torques is substantially wider in the low-mass case, and extends to inviscid discs with very low viscosity. 
At fixed $\alpha$ and $\dot{m}$, smaller MBHs have both smaller values for $r_1$ and larger values for $r_2$.

\subsection{Varying accretion rates}

\begin{figure*}
      \centering
    \includegraphics[width=0.475\textwidth]{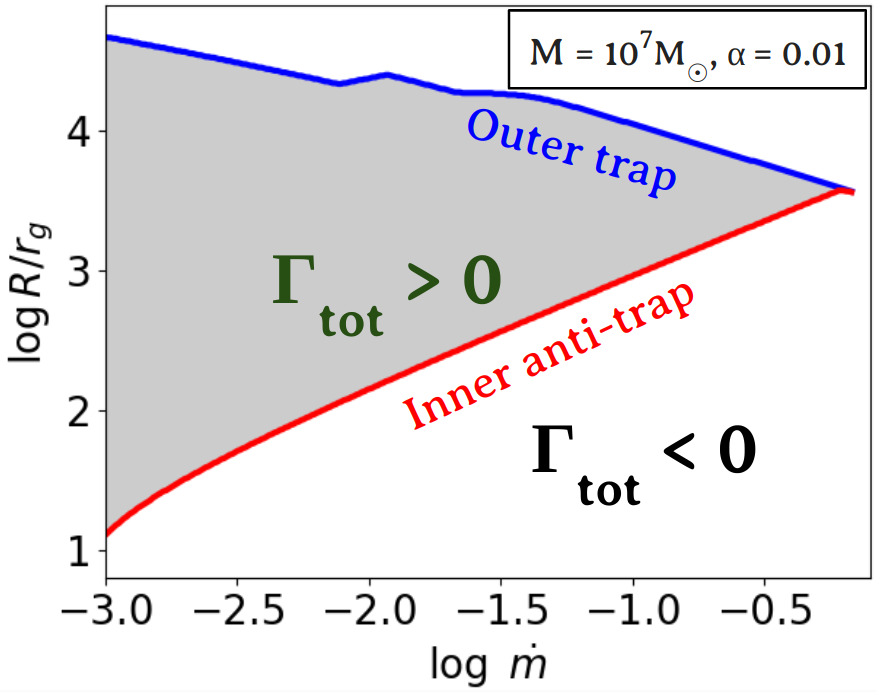}    \includegraphics[width=0.51\textwidth]{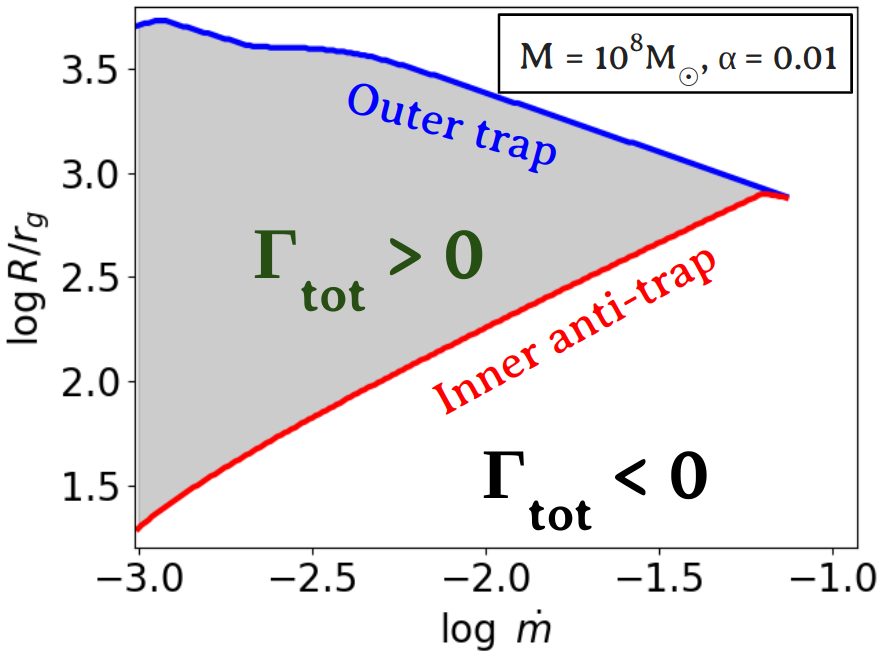}

    \caption{Zones of inward and outward migration for different accretion rates $\dot{m}$. The grey area indicate locations of positive torques, as in Figure \ref{5}. The viscosity is fixed at $\alpha=0.01$. {\it Left panel}: $M_\bullet=10^7M_\odot$.  {\it Right panel}: $M_\bullet=10^8M_\odot$.  Higher Eddington ratios eventually eliminate migration traps, particularly for high-mass MBHs.}
    \label{fig:9}
\end{figure*}

Figure \ref{fig:9} shows the same map of migration traps/anti-traps as functions of a varying accretion rate $\dot{m}$.  In these plots we fix $\alpha=0.01$ and consider $M_\bullet = 10^7M_\odot$ and $M_\bullet = 10^8 M_\odot$ cases.  Higher Eddington ratios $\dot{m}$ lead to smaller zones of outward migration, with $r_1$ increasing and $r_2$ declining.  At the highest values of $\dot{m}$, both traps and anti-traps disappear and inward migration results throughout the disc. Conversely, smaller accretion rates result in a wider area of outward migration. In the extreme cases of very inefficient accretion, $\dot{m}\lesssim 10^{-2}$, the torque is positive throughout almost the entire disc, excepting a narrow inner region dominated by GWs.



\section{Discussion} \label{5}

Our survey of migration torques in AGN discs has shown that even while updated Type I migration prescriptions \citep{jm17} do not form migration traps on their own, the inclusion of thermal torques \citep{masset17, hankla2020} will robustly generate traps and anti-traps across most (but not all) of AGN parameter space. The general pattern we find is that when a zone of outward migration exists, it is bounded on the inside (at $R=r_1$) by an anti-trap and on the outside ($R=r_2$) by a true trap. All else equal, low-mass MBH discs have larger zones of outward migration, while high-mass MBHs have smaller ones, and sometimes lack migration traps entirely (Figure. \ref{fig:6} and \ref{fig:7}). Smaller viscosity parameters $\alpha$ push the positive torque zone inward, and can also eliminate it for extremely low $\alpha$ (Figure \ref{fig:8}). A high accretion rate $\dot{m}$ shrinks the zone of positive torque, and in combination with a relatively large $M_\bullet$, can eliminates migration traps altogether.  Low accretion rates expand the domain of positive torque (Figure \ref{fig:9}).  In this section we discuss caveats to these findings as well as astrophysical implications.  

\subsection{Implications for BH populations}

\begin{figure*}
      \centering
    \includegraphics[width=0.48\textwidth]{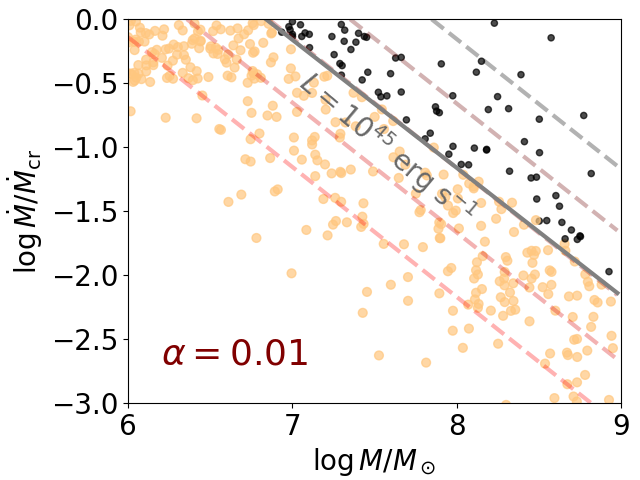} \includegraphics[width=0.48\textwidth]{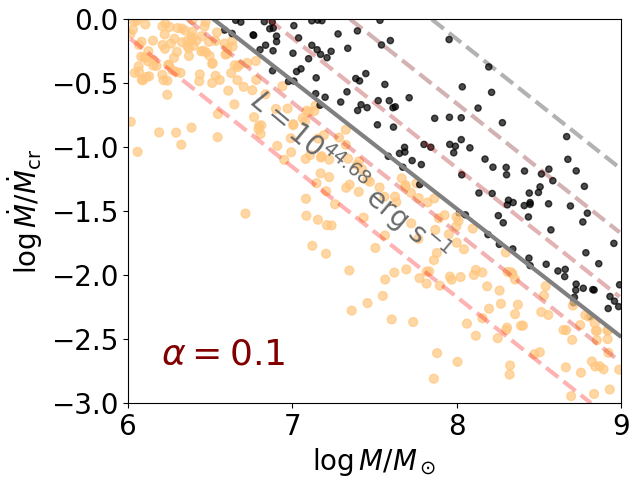}
    
    \caption{The dichotomous range of AGN migration traps in $\{M, \dot{M}\}$ space. The large orange points are discs that have traps, while the small black points are discs that do not. The dashed curves are lines of constant $L_{\rm AGN}$ from $10^{44} \ \rm erg\ s^{-1}$ to $10^{46} \ \rm erg\ s^{-1}$ in jumps of $0.5$ dex. The thick grey line is the approximate bifurcation boundary between discs that have traps and discs that do not. {\it Left panel}: $\alpha=0.01$. {\it Right panel}: $\alpha=0.1$. The highest luminosity AGN lack migration traps.}
    \label{fig:10}
\end{figure*}

The pile-up of stellar-mass BHs near a migration trap leads to strong pairwise or multi-body interactions between BHs.  A combination of multi-body chaos and hydrodynamic interactions with the AGN gas forms binaries in the trap, which may then merge \citep{bellovary16, secunda19, secunda20} and produce LVK-band GWs.  The total number of binaries synthesized in these traps depends on the size and properties of their drainage basins, i.e. the regions exterior and interior to the trap that feed single BHs into it.  In general, we find that the migration time $t_{\rm mig}$ is less than 1 Myr on both sides of the trap, so both basins will generally contribute a flux of BHs into the trap\footnote{Radii outside the Toomre instability point \citep{tqm05} may contribute single BHs that form {\it in situ} \citep{stone17}, while all radii will contribute a flux of BHs that are captured by gas drag from the pre-existing nuclear star cluster \citep{bartos17, generozov23}.}.

Obviously, in cases where migration traps do not exist (i.e. combinations of high $M_\bullet$ and high $\dot{m}$), the above dynamics will not occur, and the AGN channel will be limited to other sources of BBHs: binaries formed {\it in situ} \citep{stone17}, pre-existing binaries captured into the disc \citep{bartos17}, and binaries formed dynamically via gas dynamical friction during the migration process \citep{tagawa2020, delaurentiis2023, boekholt2023}; whether migration traps dominate the total AGN channel merger rate remains highly uncertain at this point.  But even when migration traps exist, their specific radial location can qualitatively change the outcomes of BH mergers.  

Specifically, the product of a BBH merger can acquire a velocity kick due to anisotropic GW emission \citep{non-spin-bh}.  The magnitude of the kick, $v_{\rm k}$, depends on the mass ratio, spin magnitude, and spin-orbit misalignment angles.  For slowly spinning BBHs, $v_{\rm k} \sim 100~{\rm km~s}^{-1}$, but rapidly spinning black holes can produce $v_{\rm k} > 1000~{\rm km~s}^{-1}$ \citep{cam2007}, though this requires special spin misalignment configurations, and for the case of aligned or anti-aligned spins (expected for BBHs evolving in AGN discs; \citealt{stone17}), the maximum kick velocity is in the mid-$100~{\rm km~s}^{-1}$ range. While previous works (without thermal torques) have generally found Type I migration traps at $\sim 10^{1.5-3} r_{\rm g}$, these disappear in our work (due to revised Type I migration formulae) and are replaced with thermal migration traps at larger distances of $\sim 10^{3-5.5} r_{\rm g}$. While realistic GW kicks are unlikely to completely eject merger products from the system for traps at $R \lesssim 10^5 r_{\rm g}$, the larger radii of thermal traps makes it far easier for realistic (few$\times 100~{\rm km~s}^{-1}$) kicks to excite merged BHs onto high inclination orbits.  Once placed on to a high inclination orbit, a merged BH will take substantial time to re-align into the disk, potentially longer than the AGN lifetime for distant traps \citep{fabj20,nasim23}.  Our results therefore indicate that migration traps are less favorable for hierarchical BH assembly through repeated mergers than has previously been thought.  While exploring the outcomes of disc ejection go beyond the scope of this work, our results (Figs. \ref{fig:6}, \ref{fig:7}, \ref{fig:8}, \ref{fig:9}) indicate that the radii of thermal traps can be minimized for specific AGN parameters: high Eddington ratios $\dot{m}$, low viscosities $\alpha$, and high MBH masses $M_\bullet$ (though not so high that migration traps disappear altogether).  We speculate that hierarchical BH growth (e.g. into the pair instability mass gap) may preferentially occur in these types of AGN.

Our results may also carry implications for other modes of BBH assembly and merger in AGN discs, notably the gas-assisted capture scenario \citep{tagawa2020, groebner2020}.  Slow, inward migration driven by viscous torques is a key component in BBH assembly rate estimates in this scenario: without migration, \cite{tagawa2020} find that the overall number of mergers is $\sim 20$ times smaller\footnote{See their Table 3 models 1 (canonical  -- $N=7.7\times 10^3$ mergers) and 2 (no gas migration -- $N=4\times 10^2$ mergers). }. While our study of migration from revised Type I torques and novel thermal torques does not imply that discs should be in the zero-migration limit, it does show that over large portions of parameter space, the net migration torque is not weak and negative, it is large and positive (see Figure \ref{fig:7}).  It is unclear at present how gas-assisted capture will change in the presence of rapid outwards migration, but it is plausible that this could significantly change the outcome of this process.

Finally, the presence of migration traps may serve as a bottleneck preventing the formation of wet EMRIs, at least those on prograde orbits\footnote{The subset of embedded BHs on retrograde orbits will not experience thermal or Type I migration traps \citep{secunda21}.}.  As thermal traps appear to be ubiquitous for all MBHs small enough to produce {\it LISA}-band EMRIs (Figure \ref{fig:6}), this implies that BHs forming {\it in situ} will not be able to produce wet EMRIs, and the formation rates of these low-frequency GWs will be dominated by gas capture.  The largest capture cross-sections will exist for those AGN with distant anti-trap radii $r_1$, as this minimizes the inner drainage basin into the thermal trap.  While $r_1$ is insensitive to MBH mass (Figs. \ref{fig:6}, \ref{fig:7}), it can be dramatically increased for large $\alpha$ (Figure \ref{fig:8}) or large $\dot{m}$ (Figure \ref{fig:9}).  This suggests that wet EMRI rates may be dominated by the brightest AGN in the {\it LISA}-band mass range, though this speculation needs to be examined more carefully in future work.

\subsection{Multimessenger observations}

A unique feature of the AGN channel is that LVK-band GWs from merging BBHs can be accompanied by electromagnetic counterparts, either direct or indirect.
\begin{itemize}
    \item Direct electromagnetic counterpart: a transient electromagnetic flare preceding or triggered by GW emission.  In the context of the AGN channel, this would likely take the form of an AGN flare, such as the possible \citep{graham2020} but disputed \citep{palmese2021} flare associated with GW190521.
    \item Indirect electromagnetic counterpart: a statistical association between a population of GW sources and a specific host galaxy type.  More specifically, the statistical overrepresentation of AGN in LVK sky error volumes can be searched for with a sufficient number of GW events \citep{bartos2017b, veronesi2022}, and at least one recent such search has already placed limits on the contribution of high-luminosity AGN to the total BBH merger rate \citep{ver23}.  
\end{itemize}

As we have already seen, the location and even existence of migration traps are significantly correlated with both $M_\bullet$ and $\dot{m}$.  As these parameters increase, the zone of positive thermal torques (i.e. outward migration) becomes increasingly narrow, eventually disappearing once one or both of these parameters becomes too high.  If BBH mergers in AGN primarily occur in migration traps, then high mass and high accretion rate AGN will be unable to source GWs {\it and any associated electromagnetic counterparts}.  Interestingly, current efforts to constrain the fraction $f_{\rm AGN}$ of BBH mergers originating in AGN are only able to do so down to a (high) minimum AGN luminosity $L_{\rm AGN}$ \citep{veronesi2022}.  This constraint only exists for the highest-luminosity AGN for two reasons: first, high-$L_{\rm AGN}$ AGN are intrinsically rarer and easier to exclude from LVK sky error volumes \citep{bartos2017b}; second, catalogs of high-$L_{\rm AGN}$ AGN are complete out to larger redshifts that are more competitive with the LVK GW horizon \citep{ver23}.  

In order to explore in more detail the critical AGN luminosity needed for the existence of migration traps, we sample $400$ models for each value of $\alpha=0.01, 0.1$. We draw uniformly $\log m$ from the interval $[6,9]$, and the (log of) critical accretion rate from a Gaussian distribution with $\log \dot{m}= \mathcal{N} (\mu=-1.5 - \log m; \sigma=0.6)$. If $\dot{m}$ exceeds $1$, we draw $\log \dot{m}$ from a uniform distribution from the interval $[-0.5,0]$ instead. The correlated distributions are chosen to better sample the region close to the expected boundary.

Figure \ref{fig:10} shows the dichotomy of AGN migration traps combined with lines of constant AGN luminosity, which is given by 
\begin{equation}
    L_{\rm AGN} = \dot{m} L_{\rm Edd} =  1.47 \times 10^{46} \frac{\dot{M}}{\dot{M}_{\rm cr}} \frac{M_\bullet}{10^8M_\odot}\ \rm erg\ s^{-1}. \label{L_agn}
\end{equation}
The grey lines in this figure illustrate the bifurcation boundary between discs that possess migration traps, and those that do not. Lower values of viscosity increase the AGN luminosity at the bifurcation boundary in comparison to higher viscosities. The bifurcation boundary is approximately parallel to lines of constant $L_{\rm AGN}$ over the entire sampled range of $M_\bullet$ and $\dot{M}$, which allows us to determine an approximate {\it critical AGN luminosity} $L^c_{\rm AGN}(\alpha)$.  The approximate expression for this dividing line is 
\begin{equation}
    \log_{10} L^c_{\rm AGN}(\alpha) \ {\rm[ erg \ s^{-1}]} = 45 - 0.32\log_{10}\frac{\alpha}{0.01}\label{eq:l_agn_c}
\end{equation}
For $L_{\rm AGN} \gtrsim L^c_{\rm AGN}(\alpha)$, there are no migration traps and BHs will migrate and inspiral into the MBH.  From Figure \ref{fig:10}, it appears that trap-free, sub-Eddington discs can only exist around relatively high-mass MBHs, pushing EMRI emission below the {\it LISA} band.  It is possible that super-Eddington accretion discs may lack traps even for smaller MBHs closer to the center of the {\it LISA} band, but constructing such disc models goes beyond the scope of this work.

The more immediate implication of Figure \ref{fig:10} concerns indirect searches for evidence of the AGN channel in LVK datasets. Although our models are approximate in a number of ways (see below), the existence of a maximum luminosity $L_{\rm AGN}^{\rm c}$ that permits migration traps appears robust to most of our assumptions\footnote{A key exception is that time-averaged accretion rates onto embedded BHs are Eddington-limited.}.  If future statistical searches for AGN overrepresentation in LVK error volumes finds no contribution for large $L_{\rm AGN} \gtrsim L_{\rm AGN}^{\rm c}$, but a large contribution for $L_{\rm AGN} \lesssim L_{\rm AGN}^{\rm c}$, this would not only measure the total contribution of the AGN channel to volumetric merger rates, but would also identify the precise mechanism synthesizing most of the merging BBHs in AGN.


\subsection{Caveats and neglected effects}

\label{sec:caveats}

{\bf Thermal torque saturation:} Thermal torques could be significantly reduced when the mass of the migrating BH exceeds the thermal mass \citep{guilera21} 
\begin{equation}
    \frac{m_{\rm th}}{m_{\rm BH}} = \chi \frac{c_s}{Gm_{\rm BH}} = \frac{\chi}{c_s r_B} \label{m_cr}
\end{equation}
where $r_B=Gm_{\rm BH}/c_s^2$ is the BH's Bondi radius. \cite{vrm2020}  have shown that the torque can saturate due to the non-linear flow within the Bondi radius and provided a numerical fit. These simulations are done for planetary systems where the Bondi radius is relatively small compared to the scale height. For our case, the Bondi radius is much greater than both the Hill radius and the scale height (in the regions producing thermal torque traps, the Hill radius is usually the smallest spatial scale, see Fig. \ref{fig:3}).

The simulations of \cite{vrm2020} are done in a wind-tunnel setup without any shear or self-gravity. Conversely, \cite{hankla2020} used a shearing box with zero mass, which, by definition, cannot achieve torque saturation. Thus, our regime of interest (massive perturbers, substantial velocity shear) must be probed with further shearing box simulations to estimate the exact nature of torque saturation in AGN discs.  Nevertheless, we can provide an approximate estimate of how the results may change. We stress that the estimates below are heuristic and further simulation in this regime is required to capture the exact behaviour.

The differential shear velocity at a distance $\sim H$ is $\sim c_s$, so when $H < r_{\rm B}$, the non-linearity in the accretion flow takes on a Bondi-Hoyle rather than Bondi form.  Only if $r_B \le H$ are the results of \cite{vrm2020} straightforwardly applicable. For a simple prescription approximating the transition to Bondi-Hoyle accretion, we replace in Eq. \ref{m_cr} $r_B^{-1}$ by $\min\{r_B,H\}^{-1}$.  We note that in the limit where $H \ll r_{\rm B}$, there is no longer a critical mass of the perturber: \citet{masset17} first derived the thermal mass by equating the sound-crossing time inside the Bondi sphere, $t_{\rm c}=r_{\rm B}/c_{\rm s}$, to the thermal diffusion time over the Bondi radius, $t_{\rm d}=r_{\rm B}^2 / (4\chi)$ \citep{masset_v17}.  Replacing $r_{\rm B}\to H$, we find that the condition $t_{\rm c} > t_{\rm d}$ reduces to the condition that the Prandtl number ${\rm Pr}=\nu / \chi < 4 \alpha$.  In the limit of a steady-state disk with thermal diffusivity originating from radiative diffusion (i.e. defining $\chi$ with Eq. \ref{chi-fin}), this inequality becomes $\alpha > 1/20$ (for $\gamma=5/3$), which will often be satisfied for angular momentum transport from magnetorotational turbulence.  This condition is more easily satisfied at the largest radii ($Q_{\rm T}=1$), where Eq. \ref{chi-fin} does not apply and Pr falls rapidly (Fig. \ref{fig:3}).

The numerical fits for the thermal torques are given by 
\begin{equation}
    \Gamma_{\rm th,corr} = \Gamma_{\rm hot} \frac{4\mu_{\rm th}}{1+4\mu_{\rm th}} + \Gamma_{\rm cold} \frac{2\mu_{\rm th}}{1+2\mu_{\rm th}}. \label{torque_sat}
\end{equation}
where the hot torque refers to the positive term proportional to $L/L_c$, and the cold torque refers to the negative term. Here, $\mu_{\rm th} = m_{\rm th}/m_{\rm BH}$ is the ratio between the thermal and BH mass. 

\begin{figure}
      \centering
  \includegraphics[width=0.43\textwidth]{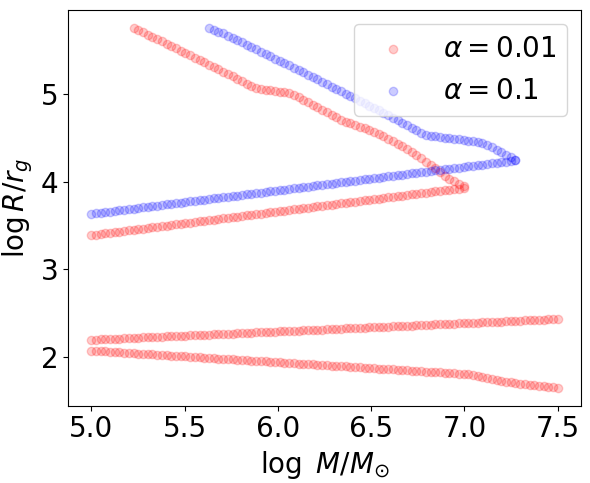}
    \includegraphics[width=0.45\textwidth]{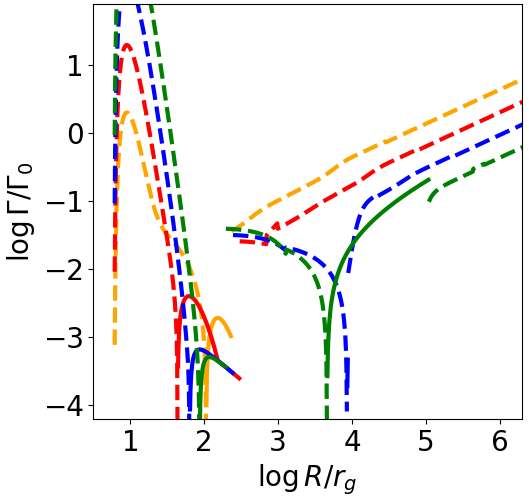}     
    
    \caption{ {\it Top panel}: Locations where the total changes sign for $\dot{m}=0.1$, similar to Fig. \ref{6}, but with thermal torque saturation from Eq. \ref{torque_sat}. Blue points for $\alpha=0.1$ models, where red points are for $\alpha=0.01$ models. {\it Bottom panel}: total torque for a variety of disc masses, similar to Fig. \ref{4}, with the same colour code.}
    \label{fig:thermal_mass}
\end{figure}

In order to explore the effects of thermal torque saturation, we modified the thermal torque according to Eq. \ref{torque_sat} and looked for the location where the total torque changes sign. The upper panel of Fig. \ref{fig:thermal_mass} shows the result for two values of $\alpha$. In both cases, the 'bifurcation point' for which migration traps vanish is achieved for slightly lower SMBH mass and larger values of $r$. While lower $\alpha$ leads to lower critical SMBH mass (due to greater reduction in the hot thermal torque), it also generates an inner migration trap around $200-300r_g$ throughout all the masses. 

The bottom panel of Fig. \ref{fig:thermal_mass} shows the total torques for different SMBH mass, where the formation of a positive torque region is found. This region forms even though both components of the thermal torque are saturated because the cold component is more saturated than the hot one, and a lower luminosity threshold is required for the net thermal torque. Solving for $\Gamma_{\rm th, corr}=0$ leads to 
\begin{equation}
    \frac{L}{L_c} = \frac{1+4\mu_{\rm th}}{2 + 4\mu_{\rm th}}
\end{equation}
such that $L/L_c\to 1$ in the unsaturated limit ($\mu_{\rm th} \gg 1$), and $L/L_c \to 1/2$ in the strongly saturated limit ($\mu_{\rm th} \ll 1$). Remarkably, $L/L_c$ is constant in zone I since $L/L_c \propto (\rho \chi)^{-1} \propto H / \Sigma \nu $ and $H$ is independent of the radius. Fig. \ref{3} shows that for our set of parameters, $L/L_c \approx 0.74$, so indeed effective torque saturation can reverse the sign of the total thermal torque.

{\bf Type II migration:} The linear theories of Type I and thermal torques both assume that the disc structure is intact, without gaps or discontinuities. However, the gravitational perturbations from a sufficiently massive perturber can overcome the viscous torques of the disc, forming an annular cavity around the massive object. The condition for gap formation is \citep{kan18}:
\begin{equation}
    q \gtrsim 5 \alpha^{1/2}h^{5/2} = 10^{-4} \alpha_{0.01}^{1/2}h_{0.033}^{5/2}.
\end{equation}
where $h=H/R$. In dimensionless form, $K\equiv q^2/\alpha h^5 \gtrsim 25$ is required to form a gap. 

These conditions are hard to satisfy for stellar mass BHs migrating in the AGN of supermassive BHs, but can be satisfied more easily if the central MBH is intermediate-mass (i.e. $M_\bullet \ll 10^6 M_\odot$; \citealt{gilbaum2022}).  
The resulting cavity has a residual surface density $\Sigma_{\rm gap}=\Sigma_0/(1+K/25)\approx \Sigma_0/K$ for large $K$. The resulting (one-sided) torque is the same as in $\Gamma_0$, but the density is replaced by the lower density $\Sigma_{\rm gap}$. Since $\Sigma_{\rm gap} \propto q^{-2}$ and the unperturbed $\Gamma_0 \propto q^2$, the modified torque is independent of the mass, which is the main feature of type II migration. 

For an embedded object massive enough to form a gap, the optical depth is also reduced by a factor of $\tau_{\rm gap} \propto K^{-1} \propto q^{-2}$, so the thermal torque may be significantly reduced. 
In general, $K\gtrsim 10^4$ is required to make the gap optically thin and dampen the torque, but more detailed studies are required. For fiducial choices of $\alpha$ and $\dot{m}$, we have seen that Type II migration is only relevant in a narrow range of $\{R, M_\bullet \}$ parameter space (Figure \ref{fig:6}, \ref{fig:7}), and outside of this our lack of Type II torques should be valid.


Type II migration may also operate in subluminous, or starved AGNs. 
Lower viscosity discs also require a lower mass threshold for Type II migration, and the gaps may be deeper under certain conditions \citep{ginzburg2018}.  Additional analytical and numerical work is required to better understand the full implications of Type II migration and subsequent BH interaction and mergers in these regimes.


{\bf Luminosity, accretion and high energy phenomena:} a key assumption in this work is that embedded BHs cannot accrete at time-averaged rates above their own Eddington limit. Without this cap, the Bondi-Hoyle accretion rate would often result in hyper-Eddington BH luminosities.  Our motivations for this cap are two-fold.  First, periods of hyper-Eddington accretion are generally thought to produce large kinetic luminosities in sub-relativistic outflows \citep{sadowski14, jiang19}.  The ram pressure of these outflows will disrupt the Bondi-Hoyle accretion flow that feeds the BH mini-disk, regulating it to a much lower time-averaged feeding rate \citep{Chen+2023}. Secondly, if even modest populations of embedded stellar-mass BHs are allowed to accrete at the (uncapped) Bondi-Hoyle rate, they would quickly consume all the gas in the AGN disk, preventing accretion onto the MBH and creating a system that does not resemble real AGN \citep{gilbaum2022}.  

However, even if the time-averaged accretion rate is Eddington-capped, brief periods of hyper-Eddington accretion may alter the local disc structure in ways that we have not modeled here, as well as produce observable signatures.  Isotropic mechanical outflows from hyper-Eddington accretion may produce powerful transients \citep{wang21}. Advection could be the dominant cooling mechanism and a complex cocoon -- jet structure may result, similar to $\gamma$ ray bursts \citep{tagawa2022}. Winds and outflows may change the global density structure and affect the net torque \citep{Gruzinov2020}.  While the response of the AGN to cyclical accretion onto embedded BHs is a complex problem requiring future study, we think it is reasonable to neglect these effects to zeroth order, as the embedded BHs are likely to spend most of their time accreting in a sub-Eddington state, during which our torque prescriptions should be accurate.


\textbf{Additional torque sources:} there are additional torque sources that one can in principle consider.  For circular prograde BH orbits, the gas dynamical friction torque \citep{Ostriker1999} is subdominant to the Type I torque by a factor of $H^2/R^2$ (when using the sub-Keplerian approximation for the gas velocity), so we have neglected it. Eccentric prograde orbits can be dominated by gas dynamical friction if the eccentricity exceeds $H/R$, but this also leads to a rapid circularisation \citep{papa2000, gri15}, so this situation cannot be sustained for a long time. Another plausible torque source is the ``headwind'' torque \citep{Pan+2021}, where the relative velocity between the stellar mass BH and the surrounding AGN gas flow potentially leads to momentum exchange between the two. The headwind torque can dominate other hydrodynamic torques if the accretion rate of the mini-disk is determined by the uncapped Bondi-Hoyle-Littleton accretion rate, but for reasons discussed above \citep{gilbaum2022, Chen+2023}, we assume that time-averaged accretion rates of embedded BHs are Eddington-capped, which  makes this torque source negligible.     Furthermore, we also note that our torque prescriptions are valid only in the thin disc limit, and do not take into account effects that may be important for $H/R \sim 1$, as suggested in \citep{PengChen2021}.  This limits the applicability of our calculations when $\dot{m} \gtrsim 1$.

\textbf{Disc response:} the potential accumulation of many BHs in a small volume that radiate near their Eddington luminosity will locally heat the gas. This type of accretion feedback may change the structure of the disc \citep{gilbaum2022}, requiring the use of more complicated disc models in regions of high BH density.  Generally, more BH heating will reduce the local optical depth, potentially deactivating the thermal torque and transforming a region of outward migration into one of (Type I dominated) inward migration, until the BHs drain to smaller radii, the AGN disc becomes optically thick again, and the cycle repeats. 
In Toomre unstable zones, supernova explosions prior to BH formation can also form cavities and observable transients \citep{grishin2021}.


\textbf{Dynamics:} we consider the evolution of a single BH migrating on a circular, prograde orbit. Type I torques tend to reduce the eccentricities and inclination of massive objects, motivating this assumption. However, thermal torques may in some circumstances increase the eccentricity and inclination of the objects \citep{masset_v17, Fro19}, an effect which needs more careful examination in future work. Furthermore, instead of a single BH, BBHs can also be present in the AGN and migrate in the disc. The differential torque on the binary elements \citep{ba11, g16} and/or the circumbinary disc formed around them may lead to a merger and varying luminosity \citep{siwek23}. Alternatively, outflows can widen and break up the binary \citep{Gruzinov2020}. We do not consider these effects in this work.  Finally, we note that even in the absence of a formal migration trap where the total torque flips sign, a ``traffic jam'' (produced by e.g. the slowdown of BHs approaching a type II gap; \citealt{tagawa2020}) or ``Zeno trap'' (produced by e.g. the slowdown of BHs in the Zone I of an increasingly dilute, radiation-pressure-dominated disc, as in Fig. \ref{fig:4}; \citealt{gilbaum2022}) may lead to a pile-up of migrating BHs, enhancing the chances for BBH formation.  We do not systematically search for these traps in our disk models, as their existence will be sensitive to the AGN lifetime.

\section{Conclusions} \label{6}

We have constructed analytical models for AGN discs with multiple zones and explored in detail the effect of thermal torques on the existence of migration traps and anti-traps (locations where the total migration torque flips sign). The radial extent of positive torques in AGN discs critically affects the rates and properties of
{\it LISA}-band extreme mass ratio inspirals and LVK-band GW mergers. We summarise our key findings as follows:

\begin{enumerate}
\item Contrary to past work, we find that Type I migration is unable to reliably produce migration traps.  More recent updates to Type I migration prescriptions, calibrated from 3D \citep{jm17} rather than 2D \citep{paardekooper2010} simulations, produce negative-definite Type I torque across the broad parameter space of AGN discs considered in this paper.  We have reached this conclusion across an exhaustive grid of relatively simple (i.e. multi-zone) disc models, and it appears to also be robust in our comparisons to a smaller number of more realistic steady-state disc models.
\item Thermal torques can dominate Type I torques over a wide range of radii and are often the dominant migration mechanism.  These thermal torques frequently take a positive sign, producing a region of outward migration that is bounded on the outside by a migration trap and on the inside by an anti-trap.  Thermal torques can become negative-definite for the highest MBH masses and Eddington ratios, in which case migration traps disappear.  The optical depth required for the thermal torque to dominate is large (Eq. \ref{tau-2}) and in fact is $\approx 2$ orders of magnitude larger than previously estimated (\citealt{hankla2020}; see \S \ref{3.4} for discussion).
\item The migration traps produced by thermal torques in our model occur at substantially larger radii than previously identified migration traps produced by older Type I migration formulae \citep{bellovary16, secunda19, gilbaum2022}, typically ranging between $\sim 10^3 r_{\rm g}$ (for large MBHs) to $\sim 10^5 r_{\rm g}$ (for the smallest supermassive BHs).  The lower Keplerian speeds at these larger radii make it easier for merging BHs to escape the traps onto inclined orbits due to GW recoil kicks, potentially delaying or altogether halting hierarchical BH growth scenarios.  
\item At the highest AGN luminosities, thermal torques become negative-definite and we find that migration traps disappear completely.  Migration traps disappear for Eddington ratios $\dot{m}\sim 1$ for MBHs as small as $M_\bullet \sim 10^7 M_\odot$, but for much lower Eddington ratios for larger MBHs (Figure \ref{fig:10}.  The critical luminosity threshold for the loss of migration traps is $L\sim 10^{44.5-45}~{\rm erg~s}^{-1}$, suggesting a natural threshold to use in future indirect searches for an AGN-GW association \citep{bartos2017b, veronesi2022, ver23}
\item For MBHs that are small enough to produce {\it LISA}-band EMRIs, sub-Eddington AGN accretion will always be accompanied by a thermal migration trap, potentially choking the supply of ``wet EMRIs'' to the central MBH.  The cross-section for gas capture of wet EMRIs is maximized in this scenario (i.e. the trap's inner drainage basin is minimized) for large Eddington ratios $\dot{m}$ and Shakura-Sunyaev viscosity parameters $\alpha$, suggesting that high-$L_{\rm AGN}$ AGN may be the most promising wet EMRI hosts.
\end{enumerate}
 
We have made a number of assumptions and idealizations in this work that deserve more careful examination in the future.  By far the most important of these is our assumption that the time-averaged luminosity of an AGN-embedded BH is near its own Eddington limit.  While we think this is a reasonable assumption (see Sec. \ref{sec:caveats}), it is not rigorously proven, and if incorrect would have major implications for migration mechanisms.  For example, the thermal torque would become positive over a much broader range of parameter space; likewise, the headwind torque, which we have neglected, could play a significant role.  Furthermore, our disc models represent approximate solutions to a more complete set of steady-state disc equations, which are themselves uncertain at large radii where star formation feedback \citep{sg03}, global torques \citep{tqm05}, and feedback from embedded BHs \citep{gilbaum2022} may alter the disc structure in ways we have not considered here.  We have only considered the evolution of BHs on circular orbits, but many BHs captured from the broader nuclear star cluster may enter on high-eccentricity orbits, or even retrograde ones \citep{secunda21}, whose evolution can be quite distinct.

The AGN channel remains one of the leading contenders for explaining LVK-band BBH merger signals.  However, the complex physics of BH-AGN interactions complicate quantitative predictions for event rates and parameters, and within the context of the AGN channel, significant uncertainties remain about exactly how BBHs are produced and driven to merger.  Multi-body interactions in migration traps are one of the most promising sites of BBH formation and GW production, particularly for repeated and hierarchical mergers, but the overall locations and even existence of these traps depends sensitively on the underlying physics that produces them: both the physics of the AGN disc and that of BH migration.  Our parameter space exploration has identified qualitative trends, such as the disappearance of such traps at high AGN luminosities, which should be further quantified in future theoretical calculations, as they offer a way to observationally test the role of migration traps (and the AGN channel more broadly) in assembling the observed population of LVK-band GW signals.

\section*{Acknowledgements}

We thank Ryosuke Hirai, Yuri Levin, Ilya Mandel, Christophe Pinte, Connar Rowan, Mor Rozner, Avi Vajpeyi, and Henry Whitehead for stimulating discussions. We further thank Andrea Derdzinski, Zoltan Haiman, Barry McKernan, Brian D. Metzger, Hiromichi Tagawa and Alessandro A. Trani for comments on the manuscript. We thank the anonymous referee for constructive feedback which greatly improved the paper. EG thanks Nicol{\'a}s Cuello for pointing out the effects of thermal mass explored in \cite{guilera21}. Part of this work was initiated during the COMPAS retreat in Dec. 2022, supported by ARC grant FT190100574 (CI: Mandel).  NCS and SG were supported by the Israel Science Foundation (Individual Research Grant 2565/19) and the Binational Science Foundation (grant Nos. 2019772 and 2020397).
\section*{Data Availability}

A script that computes the power-law structure and plots the figures and contains the data is publicly available on \texttt{GitHub} (\href{https://github.com/eugeneg88/AGN_migration}{https://github.com/eugeneg88/AGN\_migration}). The tabulated opacity disc solver is available upon reasonable request.


\bibliographystyle{aasjournal}
\bibliography{references} 

\appendix
\section{Low mass discs} \label{appendix}

\begin{figure*}
      \centering
    \includegraphics[width=0.8\textwidth]{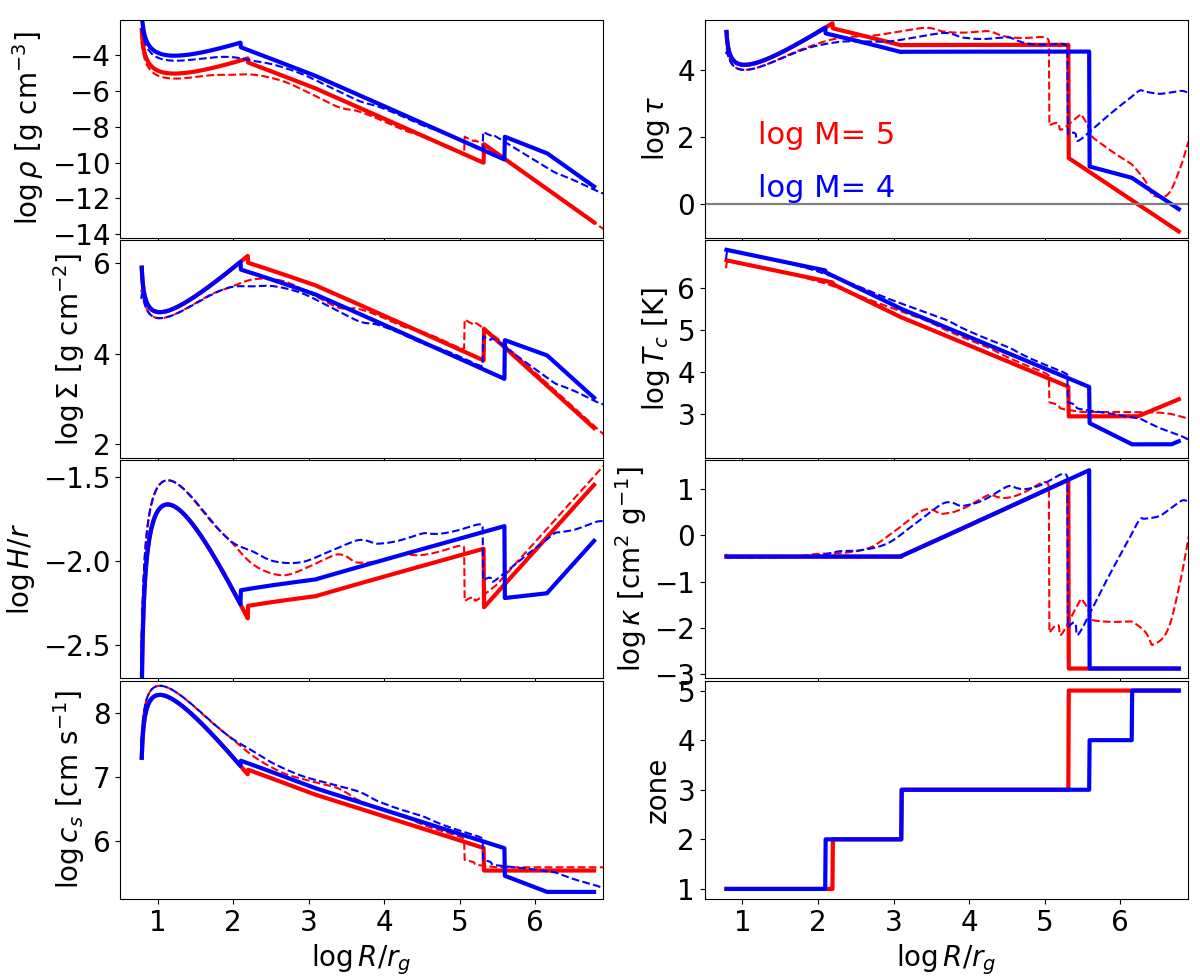}    
    \caption{The same as Figure \ref{fig:1}, but with lower MBH masses of $M_\bullet=10^5M_\odot$ (blue lines) and $M_\bullet=10^4M_\odot$ (red lines).}
    \label{fig:11}
\end{figure*}

\begin{figure*}
      \centering
    \includegraphics[width=0.73\textwidth]{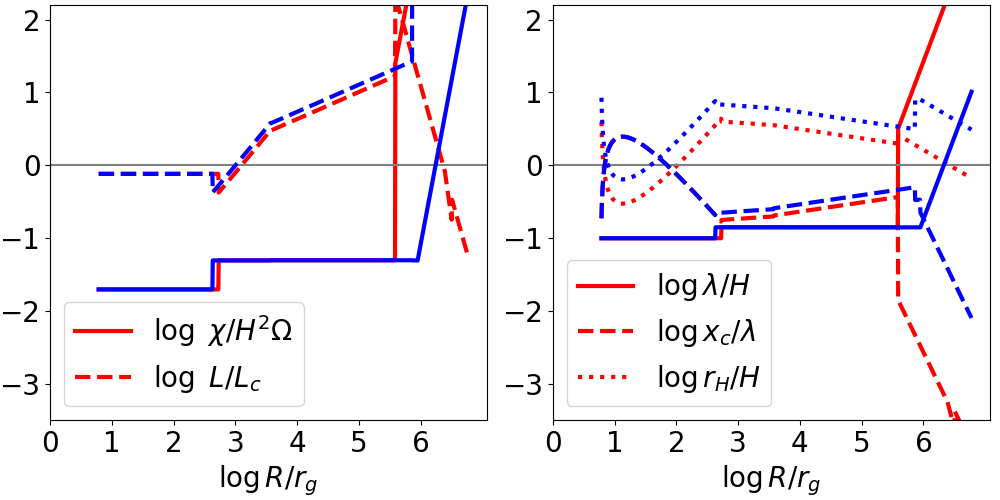}    
    
    \caption{Same as Figure \ref{fig:3}, but with intermediate-mass black holes (colors are the same as in Figure \ref{fig:11}).}
    \label{fig:12}
\end{figure*}

Figure \ref{fig:11} shows the disc solutions for intermediate-mass black holes. We see that only for the limiting case of $M_\bullet = 10^4M_\odot$, zone IV (molecular opacity) is able to form inside the transition to $Q_T=1$.  The outer zones also never become optically thin in the tabulated opacity solution due to the incorporation of dust opacities below $10^3\ \rm K$. 

Figure \ref{fig:12} shows the derived quantities for the intermediate mass black hole AGN discs. We see that in the lower mass case, neglecting self-gravity is not self-consistent, since the Hill radius $r_H$ exceeds the scale height $H$. The lower mass discs are also prone to forming a gap and entering the Type II migration regime. 
We conclude that our assumptions may be less valid in the intermediate-mass black hole regime, and that the results in this region should be examined separately in the future.


\end{document}